\documentclass[useAMS,usenatbib]{mn2e}

\usepackage{amssymb}
\usepackage{amsfonts}
\usepackage{mncite}
\usepackage{epsfig}

\newcommand{\de}{\mbox{d}}

\begin{document}

\title[Disc fragmentation during stellar encounters]{The role of the
energy equation in the fragmentation of protostellar discs during 
stellar encounters}

\author[Lodato, Meru, Clarke \& Rice]{G. Lodato$^1$, F. Meru$^1$, 
C. J. Clarke$^1$ and W. K. M. Rice$^2$\\
$^1$Institute of Astronomy, Madingley Road, Cambridge, CB3 0HA\\ 
$^2$Scottish Universities Physics Alliance (SUPA), Institute for Astronomy,
University of Edinburgh, Blackford Hill, Edinburgh EH9 3HJ}

\maketitle

\begin{abstract}

  In this paper, we use high-resolution smoothed particle
  hydrodynamics (SPH) simulations to investigate the response of a
  marginally stable self-gravitating protostellar disc to a close
  parabolic encounter with a companion discless star. Our main aim is
  to test whether close brown dwarfs or massive planets can form out
  of the fragmentation of such discs. We follow the thermal evolution
  of the disc by including the effects of heating due to compression
  and shocks and a simple prescription for cooling and find results
  that contrast with previous isothermal simulations.  In the present
  case we find that fragmentation is inhibited by the interaction, due
  to the strong effect of tidal heating, which results in a strong
  stabilization of the disc. A similar behaviour was also previously
  observed in other simulations involving discs in binary systems. As
  in the case of isolated discs, it appears that the condition for
  fragmentation ultimately depends on the cooling rate.

\end{abstract}


\section{Introduction}

Young stellar clusters are dynamic environments and
encounters between cluster members  can be quite common
(see, for example, Figure 6 of \citealt{SC01}). Such encounters
can have a significant effect on the structure and evolution of the
gaseous discs that surround young stars \citep{bate03b}. Some of these
effects include a tidal truncation or disruption of the disc
\citep{clarke93,hall96}, a burst of accretion activity onto the central
star \citep{bonnell92}, or the triggering of a gravitational
instability in the disc, which might then lead to disc fragmentation
\citep{boffin98,watkins98a,watkins98b}. The latter possibility has been
often invoked as a possible formation mechanism for low mass
companions, such as brown dwarfs, or massive planets (for some recent
reviews, see \citealt{whitworth06,goodwin06}).  However, previous
analyses of this process \citep{boffin98,watkins98a,watkins98b} were
limited by low resolution (which did not permit the proper resolution of the
disc vertical structure) and by the replacement of
the energy equation by a simple barotropic equation of state.
All these analyses, in fact, assumed
that the disc is isothermal, which might be a reasonable assumption at
very large distances (the discs considered in these works are $\approx
1000$ AU in size), but is certainly not adequate for smaller
discs. Indeed, as clearly stated in \citet{whitworth06} ``it is
important that such simulations be repeated, with a proper treatment
of the energy equation... to check whether low-mass companions can form
at closer periastra''. This is precisely the aim of the present paper.

The importance of cooling in determining the conditions under which a
gravitationally unstable disc fragments has been clearly pointed out by
\citet{gammie01} initially from local shearing-box simulations and
later confirmed in global disc simulations
\citep{rice03b,RLA05,mejia05}. The general result is that fragmentation
occurs only if the cooling timescale is smaller than a few times the
dynamical timescale in the disc. This result can be understood in the
following way. The linear stability of a rotating disc against
axisymmetric gravitational perturbations is described by the well known
Toomre parameter:
\begin{equation}
Q=\frac{c_{\rm s}\kappa}{\pi G\Sigma},
\end{equation}
where $c_{\rm s}$ is the sound speed, $\kappa$ is the epicyclic
frequency (equal to the angular velocity $\Omega$ in a Keplerian disc)
and $\Sigma$ is the disc surface density. The disc is unstable if
$Q<1$, so that local perturbations will grow on a timescale of the
order of the dynamical timescale $\Omega^{-1}$. The linear stability
then only depends on the disc temperature (through $c_{\rm s}$) and
density $\Sigma$.  However, the non-linear growth of the perturbations
heats up the equilibrium state (through compression and shocks),
increasing the value of $Q$ and hence stabilizing the disc, and
preventing fragmentation, unless the heat produced by compression can
be removed efficiently. Since the perturbation grows on the dynamical
timescale, if we want to have fragmentation, we require that
cooling acts on the same timescale.

Actually, an alternative way to destabilize the disc is to increase its
surface density (rather than decreasing its temperature), which might
result from the tidal interaction with a companion star. In a series of
papers, \citet{boffin98} and \citet{watkins98a,watkins98b} have
explored this possibility by simulating isothermal discs which are
initially stable (with $Q>1$ throughout the disc) and are destabilzed
by stellar encounters. The process appears to be successful in that in
general the disc can be effectively destabilized. However, in this
case, the subsequent evolution of the system is essentially dictated by
the isothermal condition: the perturbation will grow and produce
substellar fragments. This is indeed what is observed in these
simulations. However, it now becomes essential to determine whether - in
the presence of heating and cooling - the strong, non-linear perturbation
induced by the encounter would still be able to produce fragmentation,
or whether, similarly to the case where the perturbation is induced
more gently, through a slow cooling, fragmentation is instead
inhibited.

Similar issues have also been considered, often including the effects
of tidal heating and radiative cooling, in the context of planet
formation in binary systems. However, the results up to now appear to
be contradictory. Indeed, if on the one hand \citet{nelson2000} and
\citet{mayer05b} claim that the presence of a companion inhibits the
formation of planets via fragmentation,  \citet{boss06}, on the other hand, 
argues in favour of fragmentation. 

In this paper, we present the results of a series of smoothed particle
hydrodynamics (SPH) simulations of the encounter of a star surrounded
by a gaseous disc with a discless companion. The gas disc is here
allowed to heat up through compression and shocks and to cool down
slowly, so that in isolation the disc does not fragment but attains a
marginally stable, self-regulated state. We explore 
parameter space by varying the orbital properties of the stars, their
mass ratio, and the mass ratio between the star and the disc. {\it None of
our simulations resulted in fragmentation}, thus proving that also in
this situation the main requirement to form substellar companions is
that the cooling be fast. The disc is strongly perturbed and its
density can be enhanced significantly, but the heat provided by
compression strongly overcomes the density increase, resulting in a
substantial stabilization of the disc. 

The paper is organized as follows. In Section \ref{sec:setup} we
describe the SPH code that we use and the physical setup of our
simulations. In Section \ref{sec:reference} we describe the evolution
of our reference run. In Section \ref{sec:parameter} we discuss the
effects of changing the main parameters of the problem. In Section
\ref{sec:conclusion} we discuss our results and draw our conclusions.

\section{Numerical setup}
\label{sec:setup}

\subsection{The SPH code}

Our three-dimensional numerical simulations are carried out using SPH,
a Lagrangian hydrodynamic scheme \citep{benz90,monaghan92}. The general
implementation is very similar to \citet{LR04,LR05} and
\citet{RLA05}. The gas disc is modelled with 250,000 SPH particles
(500,000 in a run used as a convergence test) and the local fluid
properties are computed by averaging over the neighbouring
particles. The central star and the perturber are modelled as point
masses onto which gas particles can accrete if they get closer than the
accretion radius, taken to be equal to 0.5 code units for the central
star and the high mass perturber (see below) and 0.25 code units for
the small mass perturber.

The gas disc can heat up due to $pdV$ work and artificial
viscosity. The ratio of specific heats is $\gamma=5/3$. Cooling is here
implemented in a simplified way, i.e. by parameterizing the cooling
rate in terms of a cooling timescale:

\begin{equation}
\left(\frac{\de u_{\rm i}}{\de t}\right)_{\rm cool}=
-\frac{u_{\rm i}}{t_{\rm cool}},
\end{equation}
where $u_{\rm i}$ is the internal energy of a particle and the cooling
timescale $t_{\rm cool}$ is assumed to be proportional to the dynamical
timescale, $t_{\rm cool}=\beta\Omega^{-1}$. We know from previous work
\citep{gammie01,RLA05} that if $\beta< 3-7$ (depending on the ratio of
the specific heats) the disc is unstable to fragmentation. Our aim is
to establish whether a disc that would not fragment in isolation can be
driven to fragmentation as a result of a stellar encounter. We have
therefore set $\beta=7.5$, so that in isolation our disc is not
expected to fragment.

\subsection{Disc setup}

The main physical properties of the disc at the beginning of the
simulation are again similar to those of \citet{LR04} and
\citet{LR05}. The disc surface density $\Sigma$ is initially
proportional to $R^{-1}$ (where $R$ is the cylindrical radius), while
the temperature is initially proportional to $R^{-1/2}$. Given our
simplified form of the cooling function, the computations described
here are essentially scale free and can be rescaled to different disc
sizes and masses. For reference, we will assume that the unit mass
(which is the mass of the central star) is $1M_{\odot}$ and that the
unit radius is $1AU$. In these units the disc extends from $R_{\rm
in}=0.25 AU$ to $R_{\rm out}=25 AU$.  The normalization of the surface
density is generally chosen such as to have a total disc mass of
$M_{\rm disc}=0.1M_{\odot}$ (in a couple of cases, however, we have
also considered different disc masses, see below), while the
temperature normalization is chosen so as to have a minimum value of
$Q=2$, which is attained at the outer edge of the disc. This is obtained 
by setting $T\simeq 45$ K at the outer edge. 

Initially, the disc is evolved in isolation for several outer dynamical
timescales. The general features of this initial evolution is described
in detail in \citet{LR04}. The disc starts cooling down until the
vertical scale-length $H$ is reduced such that $H/R\approx M_{\rm
disc}/M_{\star}=0.1$. At this point the disc becomes Toomre unstable
and develops a spiral structure that heats up the disc and maintains it
close to marginal stability, i.e. close to $Q=1$, roughly independent
of radius. 

\begin{table*}
  \centering
  \begin{tabular}{cccccccc}
 Simulation &  $M_{\rm disc}/M_\odot$ & No gas particles 
& $M_{\rm pert}/M_\odot$ & Orbit & Direction & eccentricity 
& $r_{\rm peri}/$AU \\
    \hline
S1 &  0.1 & 250,000 & 0.1 & coplanar & prograde & 1 & 17 \\
S2 &  0.1 & 250,000 & 1.0 & coplanar & prograde & 1 & 17 \\
S3 &  0.1 & 250,000 & 0.1 & coplanar & retrograde & 1 & 17 \\
S4 &  0.1 & 250,000 & 0.1 & `head on' & --- & 1 & 17 \\
S5 &  0.1 & 250,000 & 0.1 & coplanar & prograde & 7 & 17 \\
S6 &  0.05 & 250,000 & 0.1 & coplanar & prograde & 1 & 17 \\
S7 &  0.2 & 250,000 & 0.1 & coplanar & prograde & 1 & 17 \\
S8 &  0.1 & 250,000 & 0.1 & coplanar & prograde & 1 & 30 \\
S9 &  0.1 & 250,000 & 0.1 & coplanar & prograde & 1 & 40 \\
H1 &  0.1 & 500,000 & 0.1 & coplanar & prograde & 1 & 17 \\
    \hline
  \end{tabular}
  \caption{\small Summary of the main physical properties of the simulations
  carried out.}
  \label{tab:table}
\end{table*}

\subsection{Adding the perturber}

Once the disc has reached the quasi-steady, self-regulated state of
marginal stability, we add the perturber star. This is initially put at
a distance of $100AU$ from the primary star, and is placed in a
parabolic orbit. At this distance, the sudden addition of the perturber
does not cause any significant perturbation to the disc structure. We
have performed a large number of simulations, by varying the orbital
parameters, considering coplanar encounters (both prograde and
retrograde with respect to the disc rotation) and `head on' encounters,
where the orbital plane of the secondary is perpendicular to the disc
plane. We have also performed a simulation of a hyperbolic (coplanar)
encounter. We have varied the disc mass, the perturber mass $M_{\rm
pert}$ and the pericenter of the perturber orbit $r_{\rm peri}$ and we
have also run simulations with different numerical resolutions. Table
\ref{tab:table} summarizes the main physical setup of the various
simulations that we have done.

\subsection{Resolution issues}

One important aspect that needs to be taken into account is whether our
simulations have enough resolution to resolve the fragments, if
fragmentation occurs. As shown by \citet{bate97}, SPH correctly
reproduces fragmentation if the Jeans mass is larger than the minimum
resovable mass, defined as:

\begin{equation}
M_{\rm min}=2N_{\rm neigh}m_{\rm i}=2M_{\rm tot}\left(\frac{N_{\rm
neigh}}{N_{\rm tot}}\right),
\end{equation}
where $m_{\rm i}$ is the mass of a single SPH particle, $M_{\rm tot}$
($=M_{\rm disc}$ in this case) is the total mass of the gas being
simulated, $N_{\rm tot}$ is the total number of SPH particles used, and
$N_{\rm neigh}$ is the number of SPH neighbours (which is always kept
close to 50). 

To estimate the expected Jeans mass $M_{\rm J}$, we use the fact that the most
unstable wavelength to gravitational instability is of the order of the
disc thickness $H$. We then have $M_{\rm J}\approx\Sigma H^2\approx
M_{\rm disc}(H/R)^2$, from which we get:

\begin{equation}
\frac{M_{\rm J}}{M_{\rm
min}}\approx
\frac{1}{2}\left(\frac{H}{R}\right)^2\frac{N_{\rm
tot}}{N_{\rm neigh}}
\approx 
\frac{1}{2}\left(\frac{M_{\rm disc}}{M_{\star}}\right)^2\frac{N_{\rm
tot}}{N_{\rm neigh}}.
\end{equation}

For most of our simulations, we use 250,000 particles, which would
then lead to $M_{\rm J}\approx 25M_{\rm min}$.  We thus have more than
enough resolution to properly address the issue of
fragmentation. However, to be sure, we have also performed a
convergence test of our results, by running a simulation with twice as
many particles. The results of this convergence test are described
below in Section \ref{sec:parameter}.

The second aspect related to resolution is that we require artificial
viscosity to play a role only when modeling shocks. In order to ensure
this, we then require that the velocity difference accross a smoothing
kernel is subsonic, i.e. $h\Omega<c_{\rm s}$, where $h$ is the
smoothing length. This in turn requires that the smoothing length is
smaller than the disc thickness $H=c_{\rm s}/\Omega$. We have indeed
checked that, even at the lower resolution of 250,000 particles, the
average smoothing length is a fraction $\approx 0.4$ of the disc
thickness. 

\begin{figure*}
\centerline{\epsfig{figure=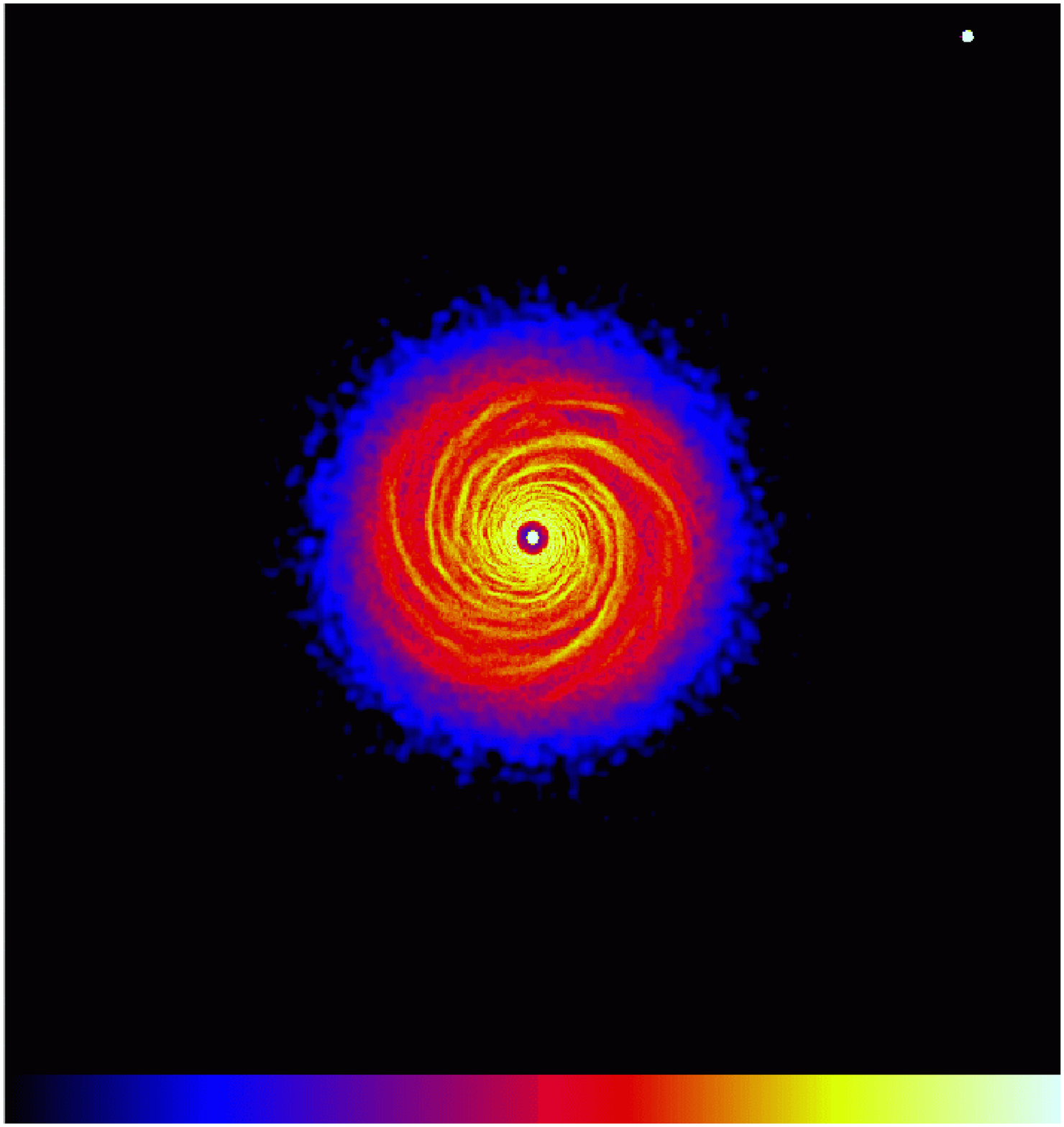,width=0.4\textwidth}
            \epsfig{figure=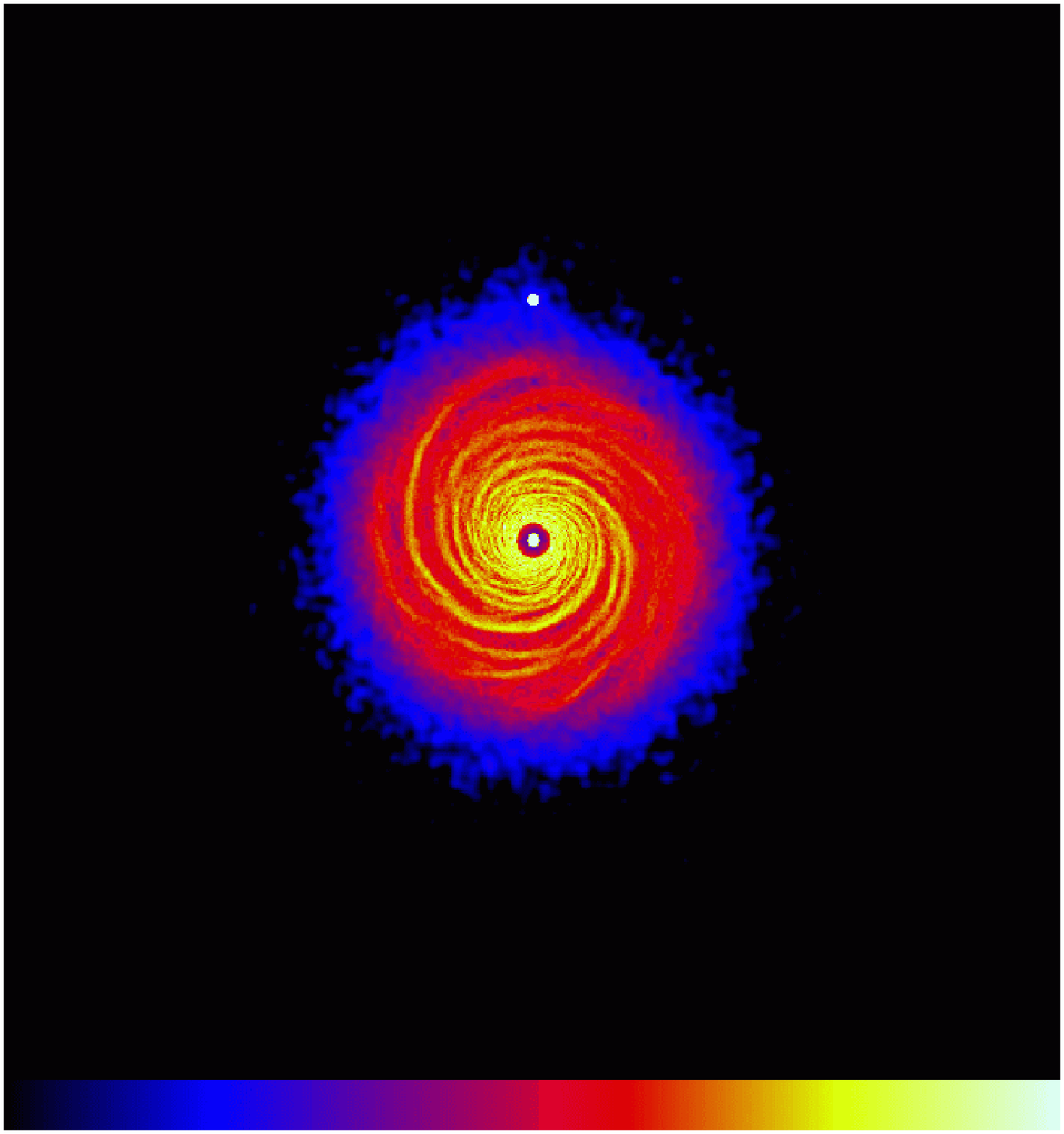,width=0.4\textwidth}}
\centerline{\epsfig{figure=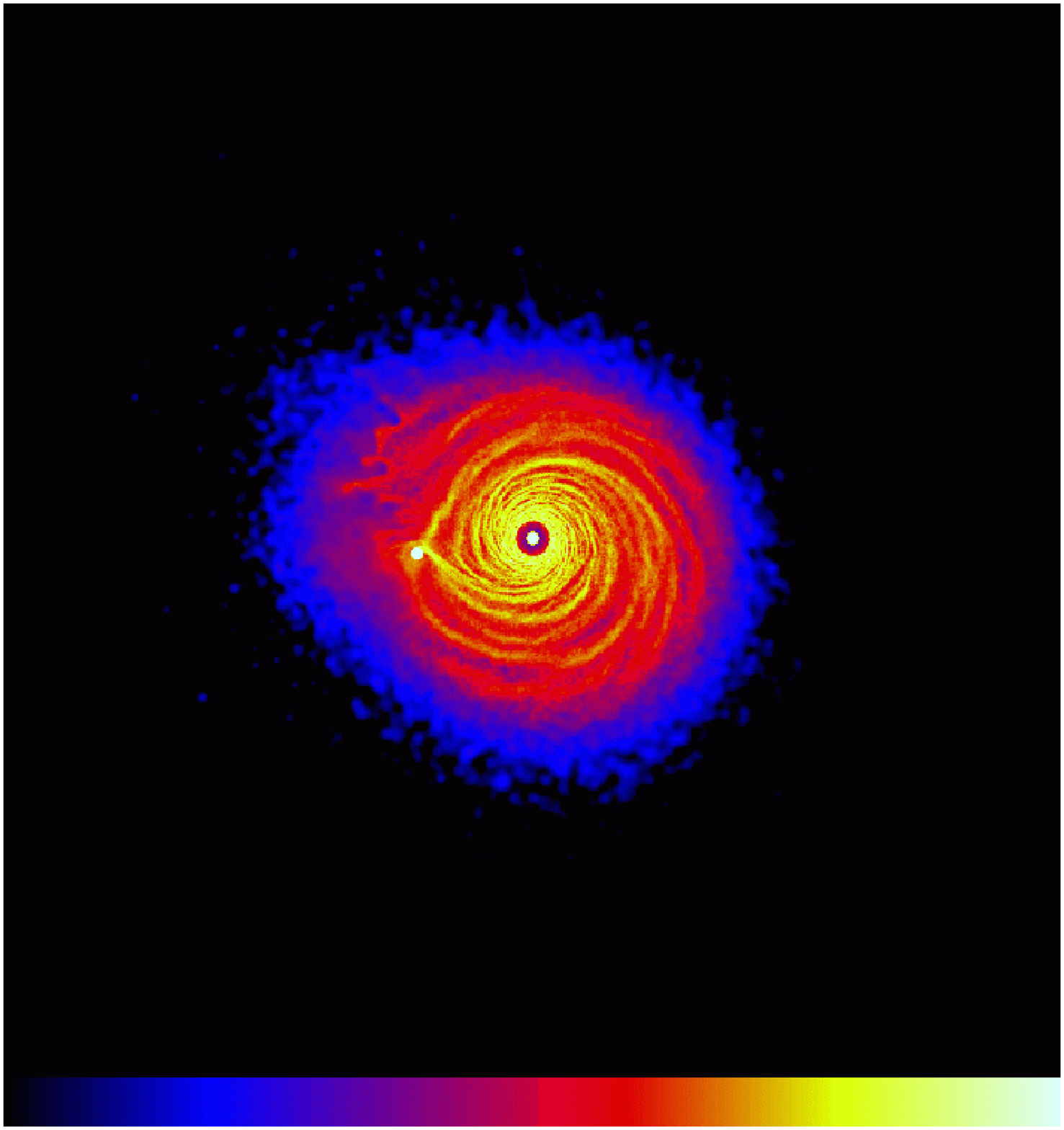,width=0.4\textwidth}
            \epsfig{figure=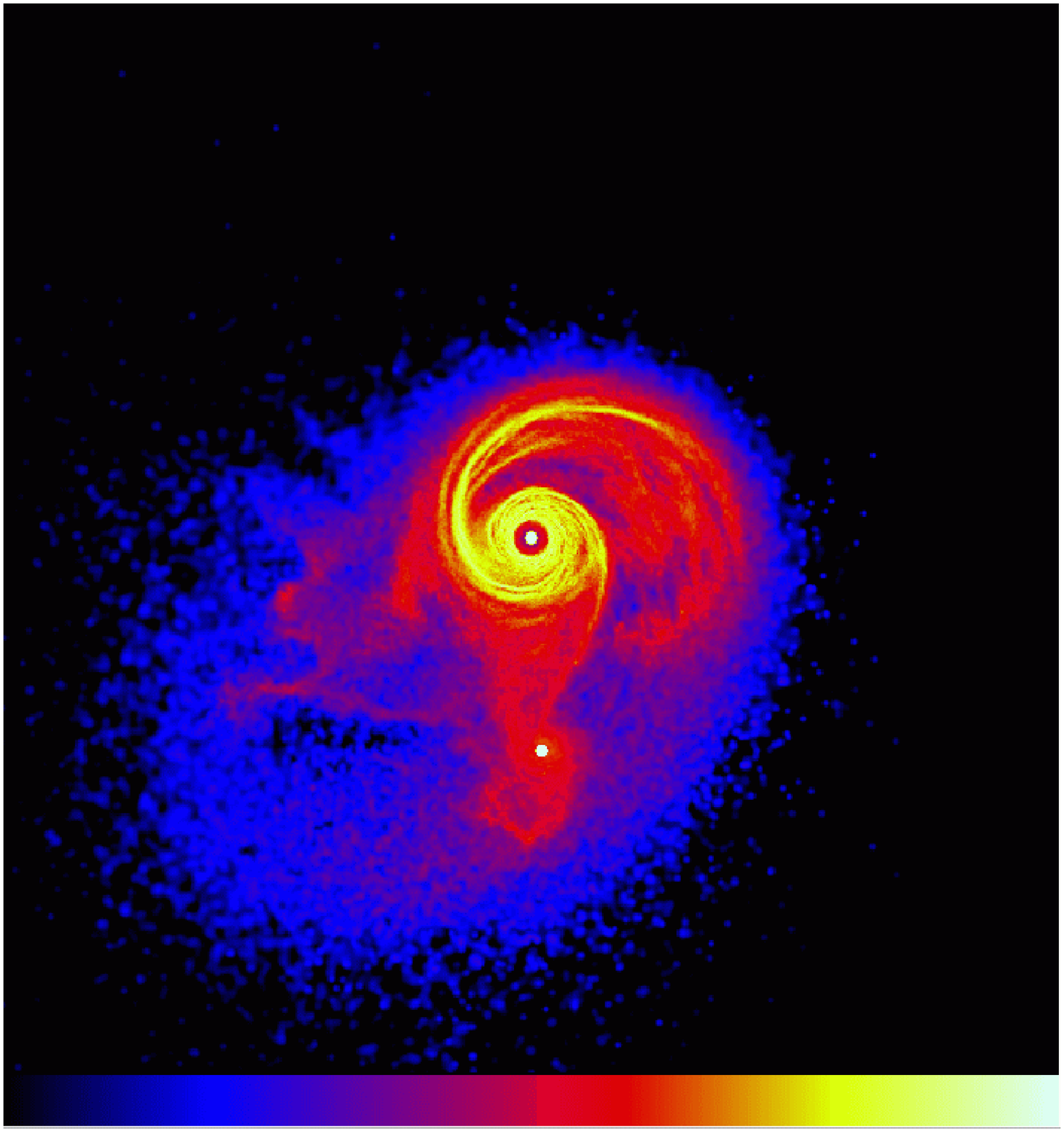,width=0.4\textwidth}}
\centerline{\epsfig{figure=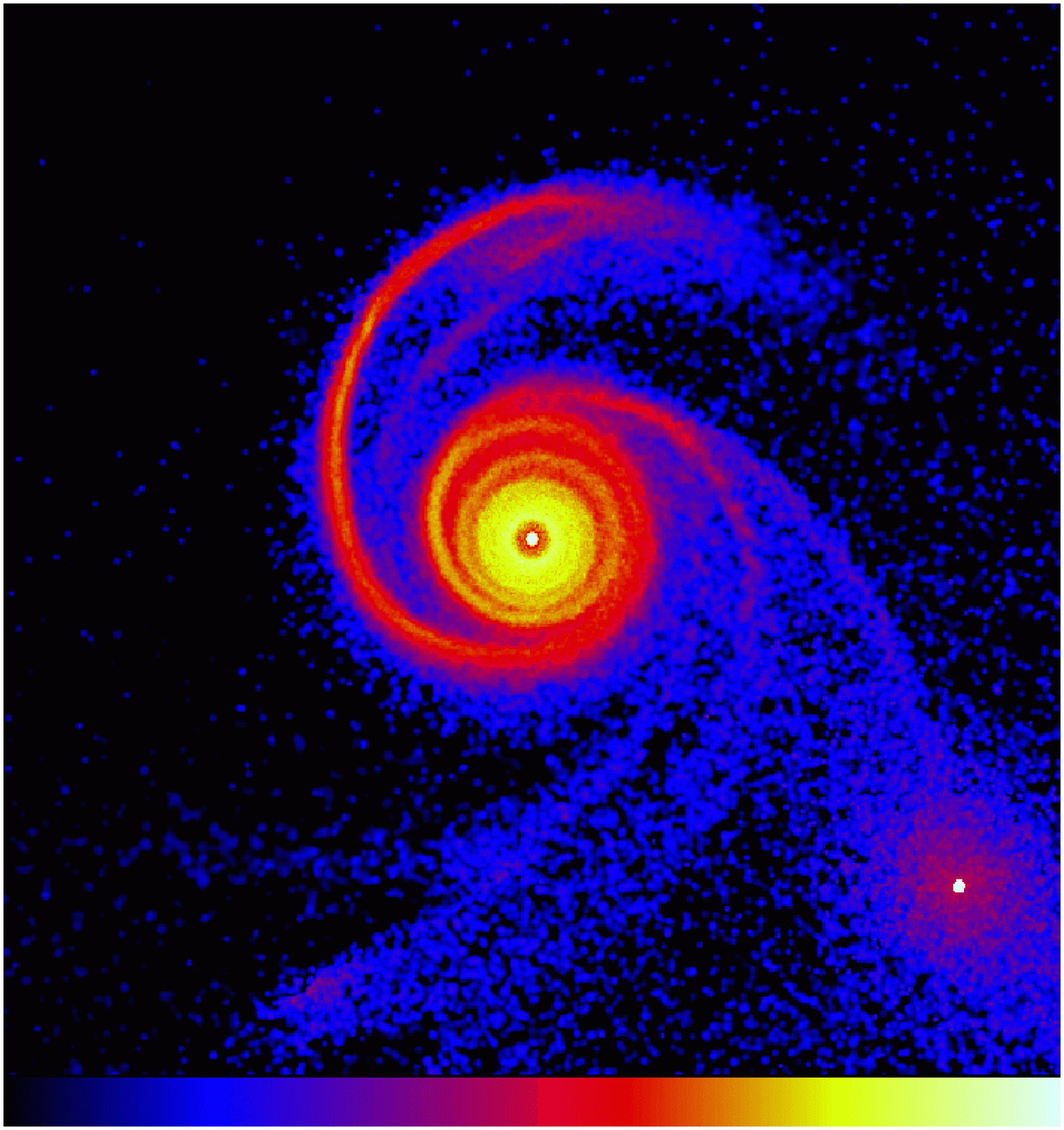,width=0.4\textwidth}
            \epsfig{figure=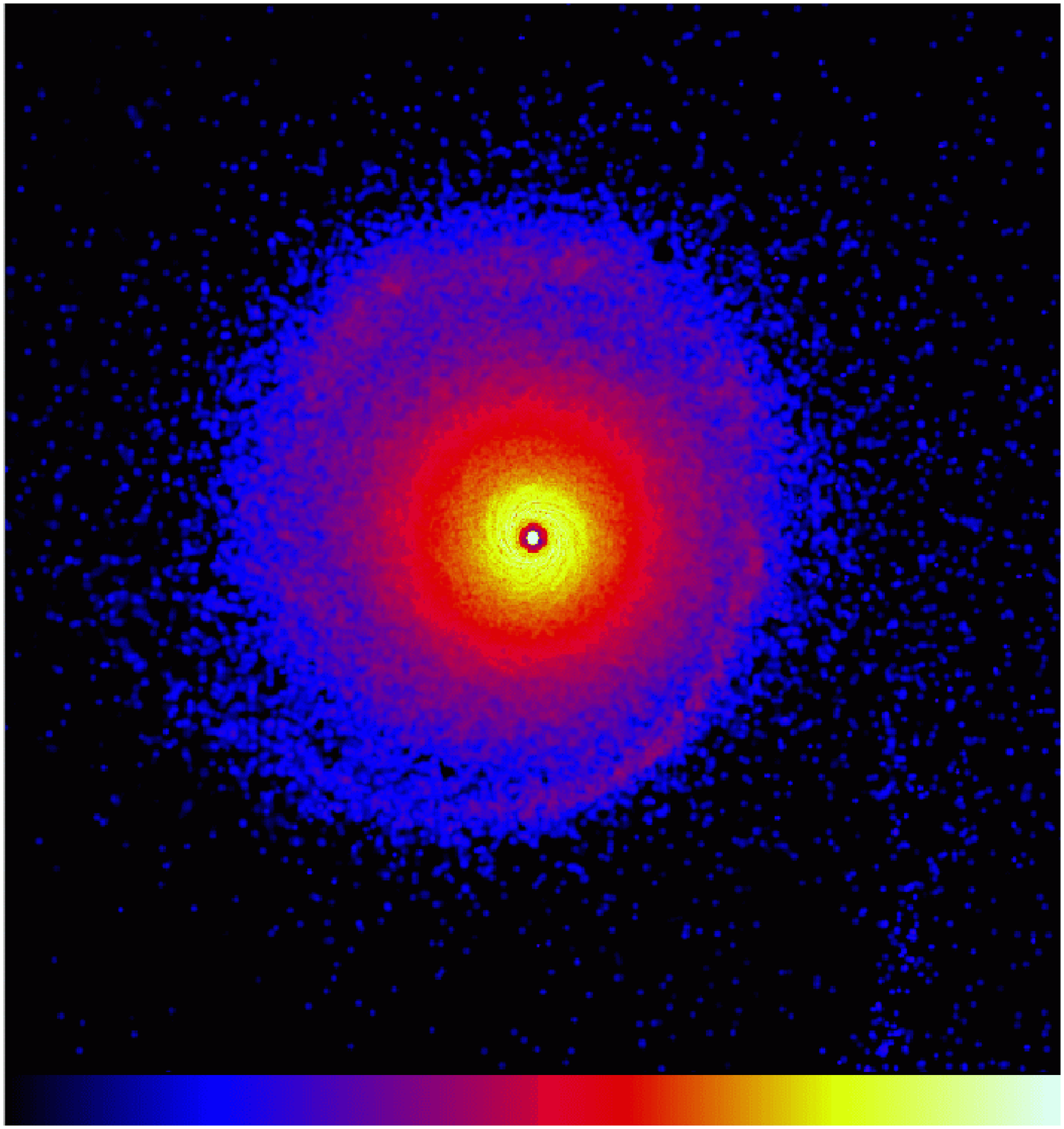,width=0.4\textwidth}}
\caption{\small Images of the reference simulation at (from top left to
  bottom right) (a) $t\simeq-90.7$ yrs (b) $t\simeq-22.3$ yrs (c)
  $t=0$ yrs (d) $t\simeq19.1$ yrs (e) $t\simeq81.2$ yrs (f)
  $t\simeq302.4$ yrs}
\label{fig:sims}
\end{figure*}

\section{The reference run}
\label{sec:reference}

In this section we describe the results of our reference simulation,
called S1, for which the main parameters are listed in the first line
in Table \ref{tab:table}. Fig. \ref{fig:sims} shows images of the disc
structure at different times during the evolution, where $t=0$
indicates the time at which the perturber reaches pericenter. The
colour scale indicates the logarithm of disc surface density between
$1~ \mbox{g}/\mbox{cm}^2$ and $10^4~\mbox{g}/\mbox{cm}^2$. The linear
size of the images is 160 AU and the images are centered on the
instantaneous position of the primary. The two stars are visible as
white dots in this image. The perturbing star is initially in the top
right hand corner of panel a.  It can be seen that the disc starts off
in a marginally stable state and displays a clearly visible spiral
structure at $t\simeq-90.7$ years (panel a).  The disc orbits in an
anti-clockwise motion.  At $t\simeq-22.3$ (panel b), the perturber
begins to interact with the disc. At $t=0$ (panel c), the perturber
reaches the pericentre. After pericenter passage, the disc structure is
significantly perturbed, as shown at $t\simeq19.1$ years (panel d).  At
$t\simeq81.2$ years (panel e), there is barely any evidence of the
pre-encounter spiral structure and the only visible structure is a
pronounced tidal tail.  The perturber has also carried away roughly 4\%
of the primary disc mass, most of it being accreted on the perturber
and the remainder forming a small circum-secondary disc, as well as
causing some gas ($\approx 6$ \% of the disc mass) to be scattered away
from both stars.  At $t\simeq302.4$ years (panel~f), the spiral
structure is completely gone leaving a highly stable larger disc in
which fragmentation has been prevented.

\begin{figure}
\centerline{\epsfig{figure=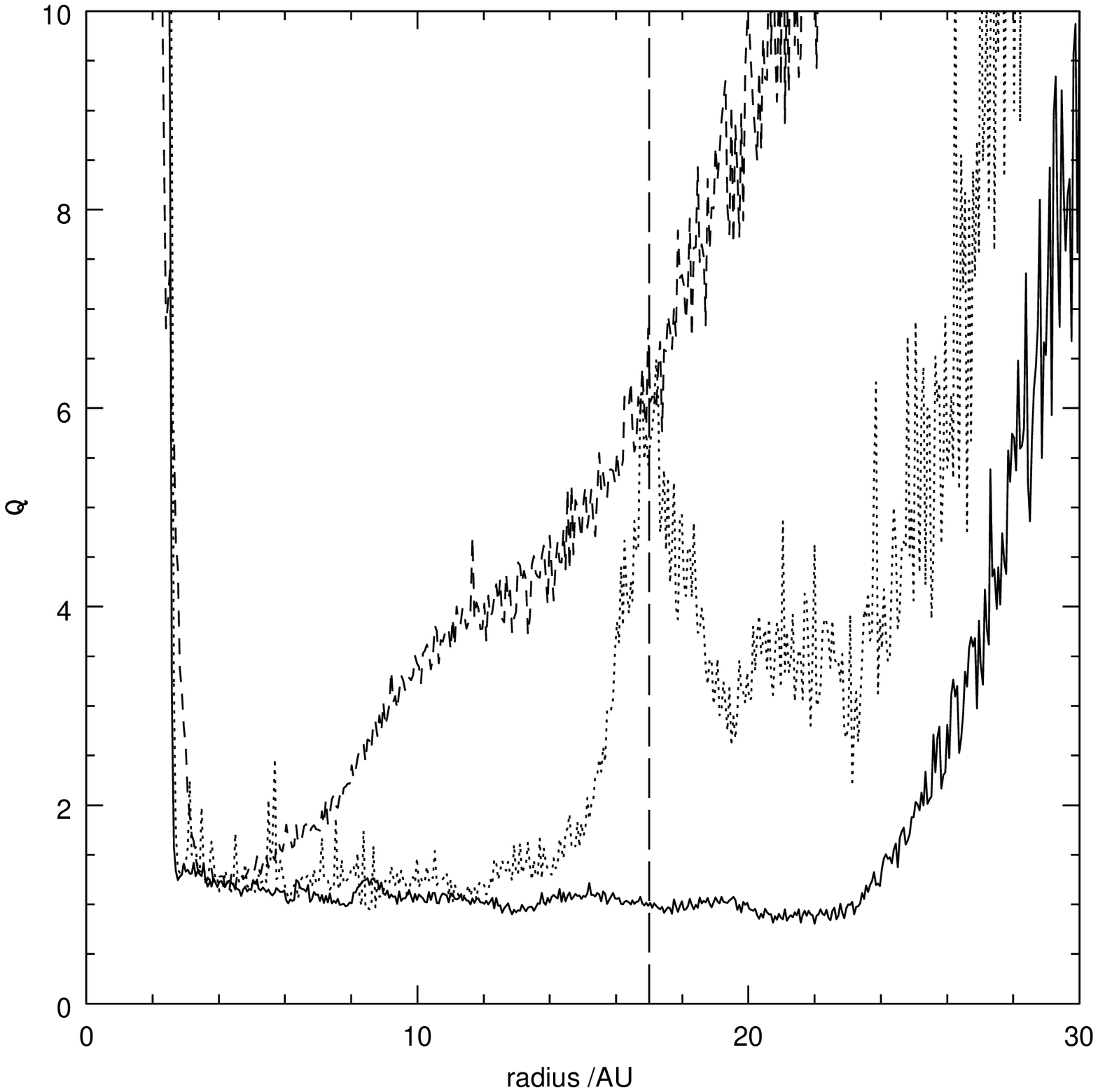,width=0.5\textwidth}}
  \caption{\small Radial profile of $Q$ at $t\simeq -90.7$ yrs (solid
  line), $t\simeq 0$ yrs (dotted line) and $t=302.4$ yrs (dashed line).
  The vertical dashed line shows the pericentre distance.}
  \label{fig:q_ref}
\end{figure}

\begin{figure}
  \centerline{\epsfig{figure=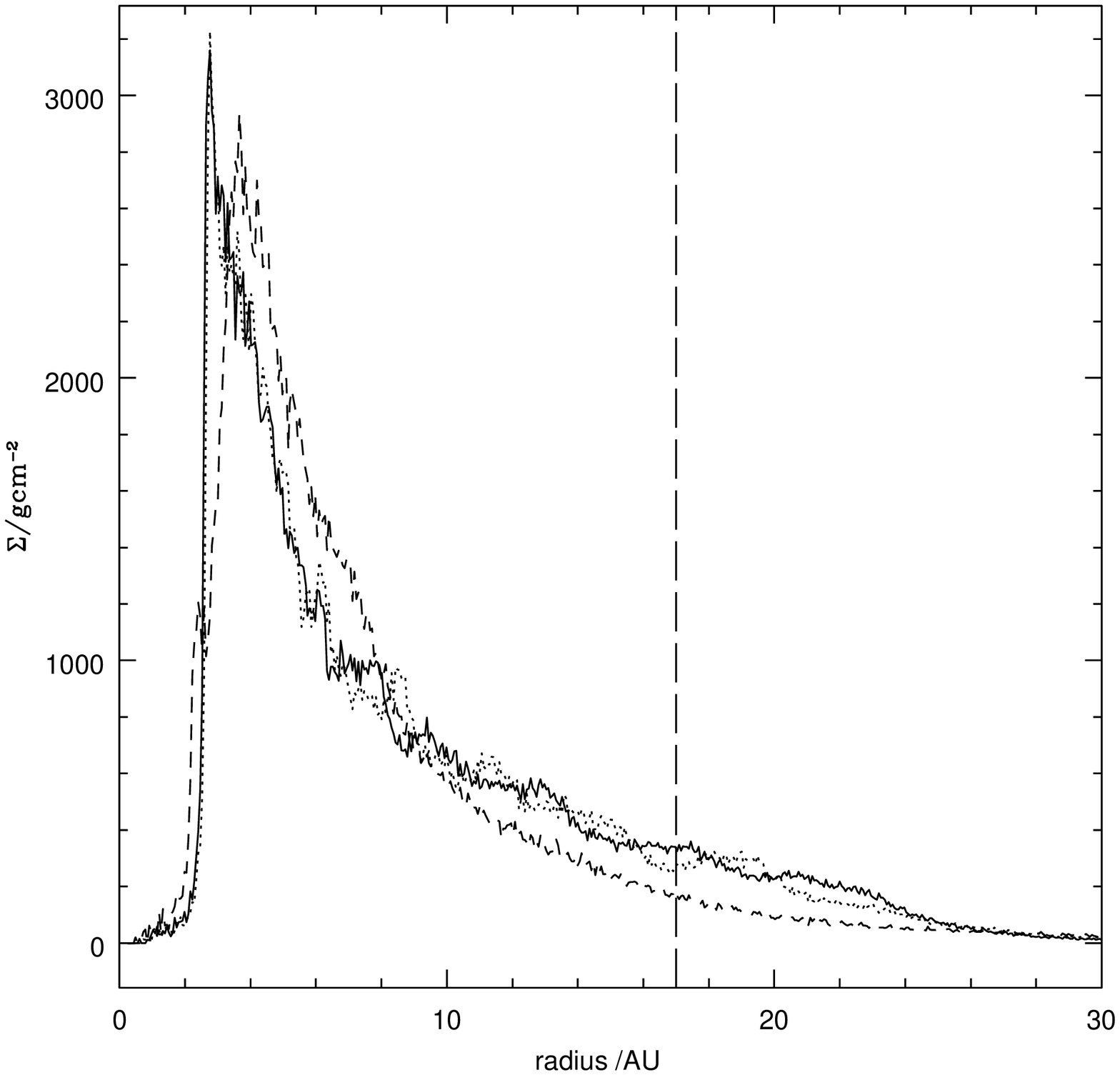,width=0.5\textwidth}}
  \caption{\small Radial profile of $\Sigma$ at $t\simeq -90.7$ yrs
  (solid line), $t\simeq 0$ yrs (dotted line) and $t=302.4$ yrs (dashed
  line).  The vertical dashed line shows the pericentre distance.}
  \label{fig:sigma_ref}
\end{figure}

Figure \ref{fig:q_ref} shows the radial profile of $Q$ for the
reference case for three different times: the start (solid) and end
(dashed) of the simulation, and the time when the perturber reaches the
pericentre (dotted).  It can be seen that when the perturber reaches the
pericentre it has stabilized roughly the whole outer disc.  By the end
of the simulation, the disc has become stable at all radii. Figure
\ref{fig:sigma_ref} shows the corresponding plots for the surface
density, $\Sigma$.  There is not much evolution in the surface density
before the perturber reaches the pericenter.  However, it can be seen
that $\Sigma$ has become significantly steeper at the end of the
simulation, suggesting an increase in the accretion rate.

\begin{figure*}
\centerline{\epsfig{figure=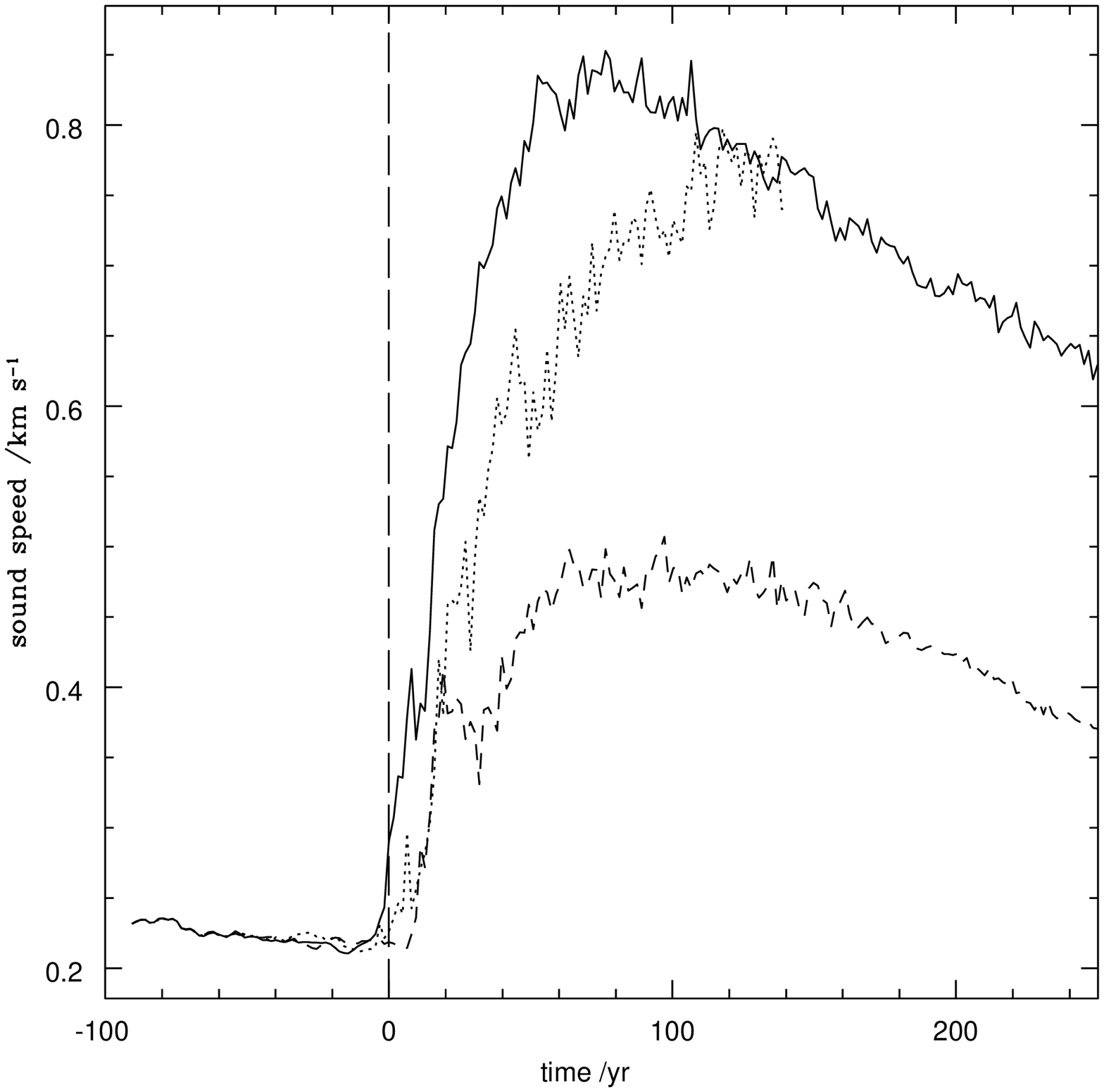,width=0.4\textwidth}
	    \epsfig{figure=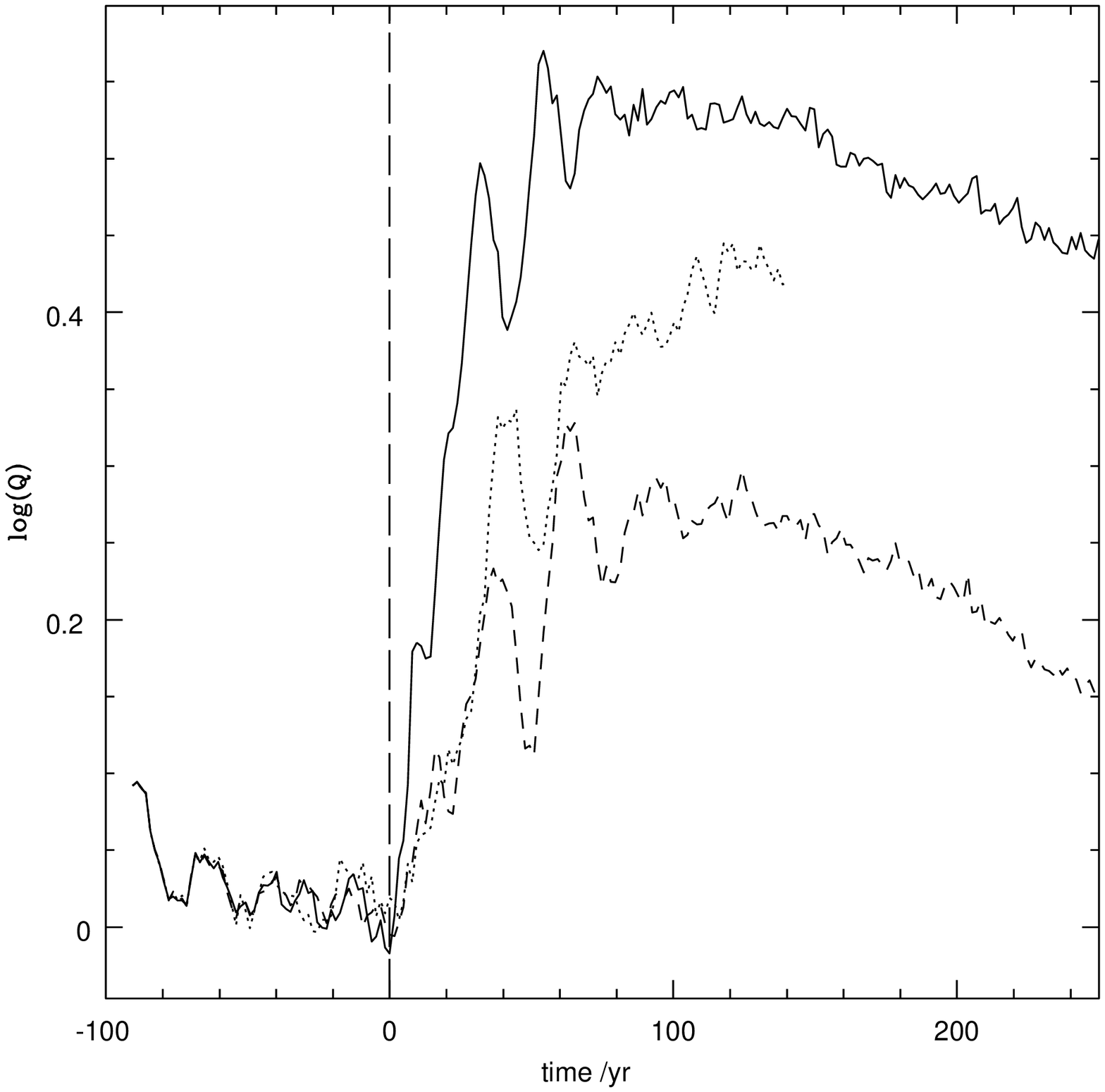,width=0.4\textwidth}}
\caption{\small  Azimuthal averages at
$R=8.5$AU of sound speed (left) and  $\log(Q)$ (right)
as a function of time
for the simulations involving prograde (solid line), retrograde
(dotted line) and non-coplanar (dashed line) orbits.}
\label{fig:angl}
\end{figure*}

A more detailed view of the effects of the interaction on the disc
structure can be obtained from Fig. \ref{fig:angl} (solid line), where
we plot the evolution of the sound speed and $Q$ at a radius $R_{\rm
  r}=8.5$ AU, i.e. halfway between the pericenter and the primary
star; the moment at which the perturber reaches pericenter is marked
with a dashed line. Several things can be noted from these
plots. First of all, the interaction produces a strong increase in
sound speed, with the local temperature increasing from about $10$ K
to more than $100$ K within a few decades of pericentre.  Thereafter
the sound speed declines on the local thermal timescale. The right
hand panel shows that $Q$ tracks the sound speed almost exactly, which
is consistent with the fact that changes in $\Sigma$ at this radius
are minor (a few tens of per cent at most). Evidently, the interaction
has increased the specific entropy of the material, which must have
been mediated by shock heating. Clearly, the net effect of the
interaction is to increase $Q$ and thus to stabilise the disc.

\section{Exploring the parameter space}
\label{sec:parameter}

\subsection{Changing the direction of the perturber's orbit}
\label{sec:direction}

Figure \ref{fig:angl} shows a comparison of sound speed and  
$Q$  at $R=8.5AU$ for the prograde, retrograde
and non-coplanar cases (simulations, S1, S3, and S4 respectively).  The
non-coplanar orbit is set up in such a way so that it impacts the disc
at $90^\circ$. It can be seen that the largest effect on the disc
structure is obtained with a prograde passage, while the retrograde and
especially the `head on' encounter have a substantially reduced
effect. The fact that the `head on' encounter has a smaller effect can
be easily understood since the timescale over which the perturber
interacts with the disc is significantly reduced (a similar effect
occurs also in the case of a hyperbolic encounter, see below). The
comparison between the two coplanar encounters (prograde and
retrograde) is in agreement with the results of \citet{hall96}, who
have shown that retrograde encounters induce a smaller amount of
angular momentum transfer between the perturber and the disc. 
We also note that the timescale on which the temperature reaches
its peak is roughly double that for the prograde encounter.

\begin{figure*}
\centerline{\epsfig{figure=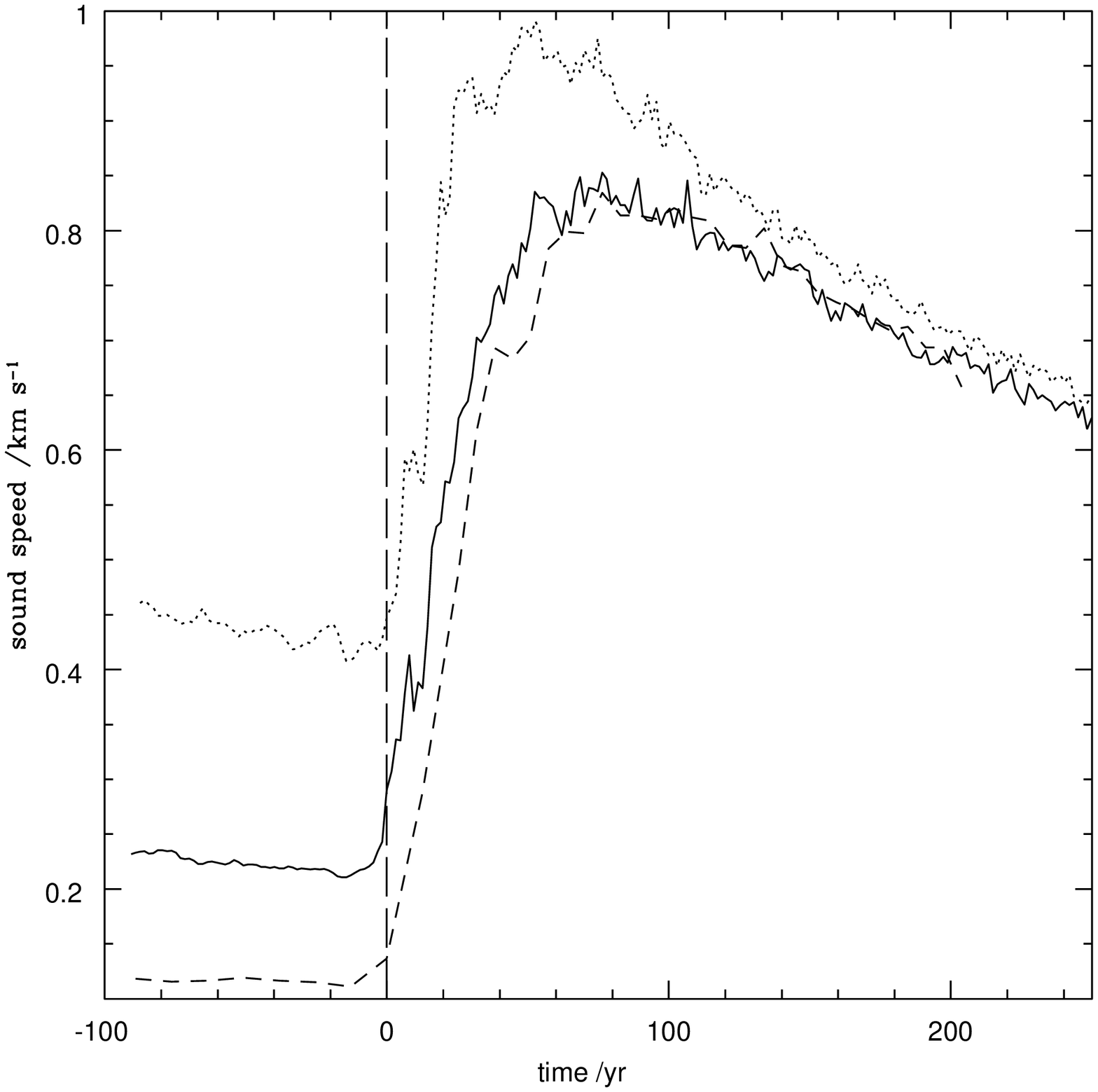,width=0.4\textwidth}
	    \epsfig{figure=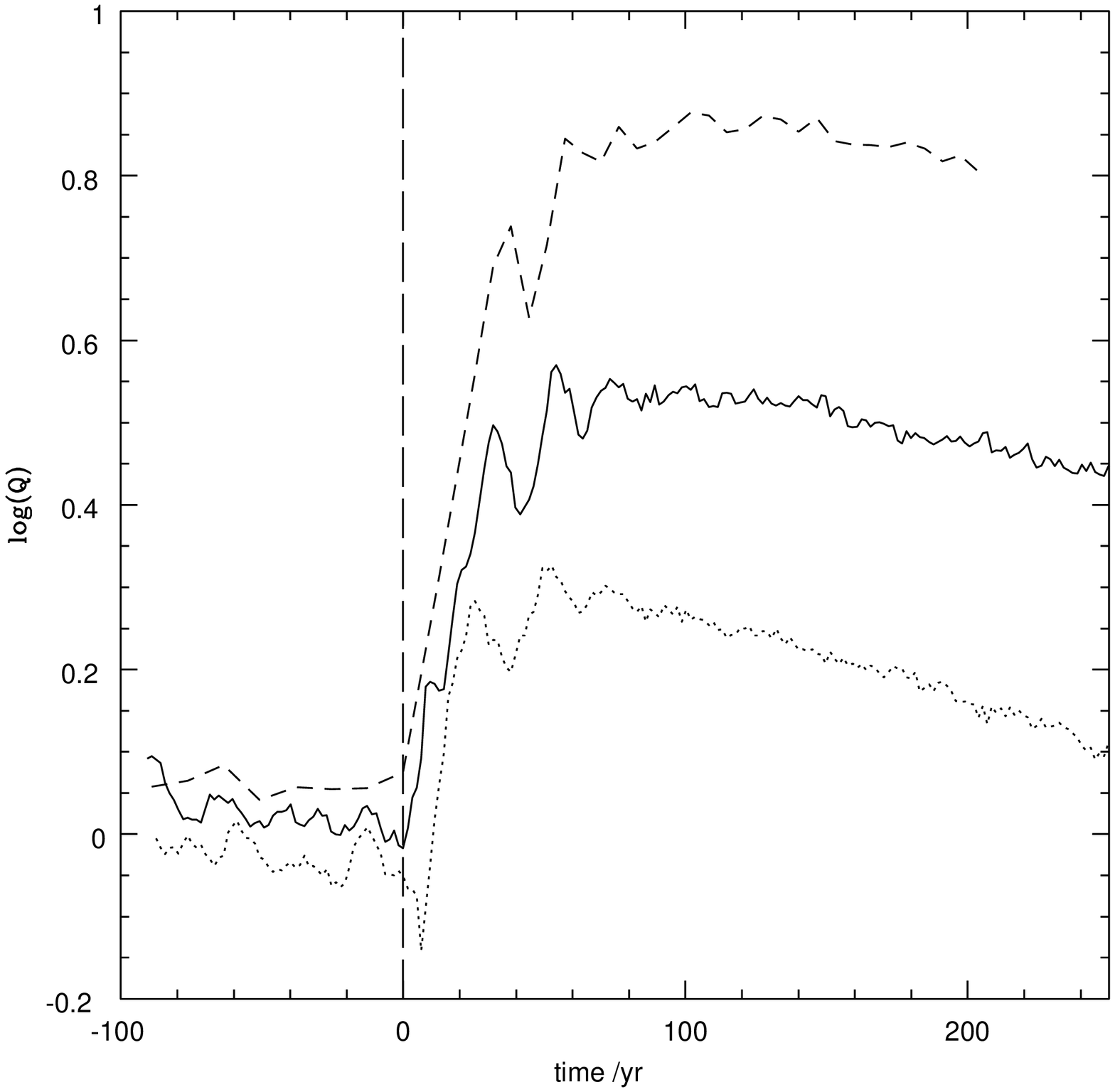,width=0.4\textwidth}}
\caption{\small  Azimuthal averages at
$R=8.5$AU of sound speed (left) and  $\log(Q)$ (right) 
as a function of time
for simulations S1 (solid line; $M_{\rm disc}=0.1M_\odot$), S7
(dashed line; $M_{\rm disc}=0.05M_\odot$) and S6 (dotted lne; $M_{\rm
disc}=0.2M_\odot$).}
\label{fig:mass}
\end{figure*}

\subsection{Changing the disc mass}

Figure \ref{fig:mass}  compares the results of simulations S1, S6 and
S7, in which we change the disc mass.  
The sound speed  profiles show
something interesting: even though the temperature of the higher mass
disc is overall the largest, its increase with respect to the
pre-encounter value is the lowest, while it is the largest for the
lighter disc.  This also shows up in the plots of $Q$, where for the
high disc mass case, the increase in $Q$ after the encounter is smaller
than for the lower disc masses.

\citet{mayer05b}, in their analysis of gravitational instability in
binary systems suggested that a lighter disc is more likely to
fragment when perturbed by a binary companion as the heating caused by
shocks is not enough to balance the large increase in density, thus
resulting in a less stable disc. However, the present simulations show
that a lower disc mass is stabilized more efficiently than a heavier
disc.  It appears that the smaller fractional temperature increase for
a high disc mass, with comparable effect on $\Sigma$, makes a higher
disc mass relatively more susceptible to fragmentation. Nevertheless,
we emphasise that for {\it all} disc masses studied, $Q$ increases
during the encounter, i.e. the effect of the encounter is one of
stabilisation.

\begin{figure*}
\centerline{\epsfig{figure=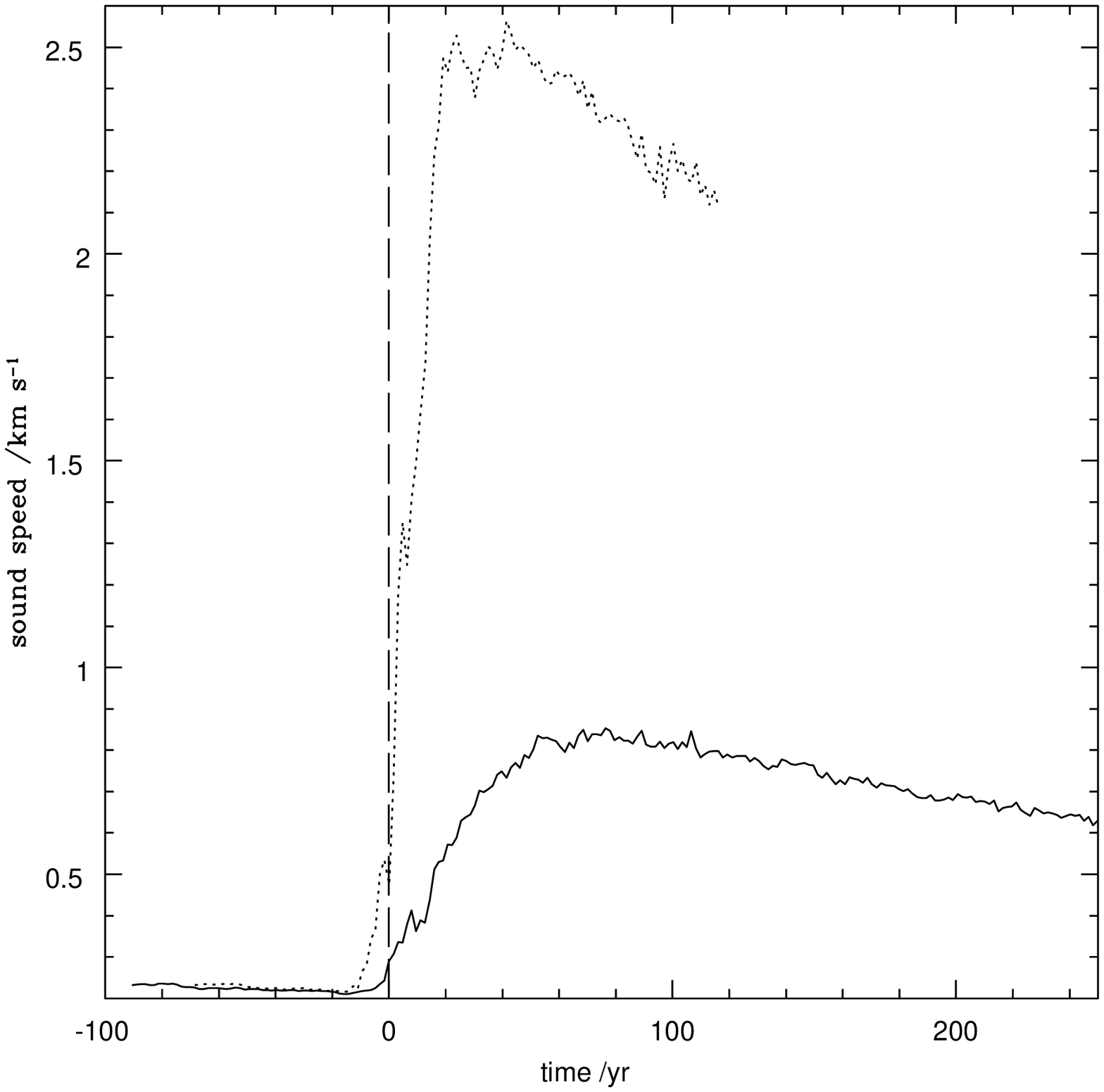,width=0.4\textwidth}
	    \epsfig{figure=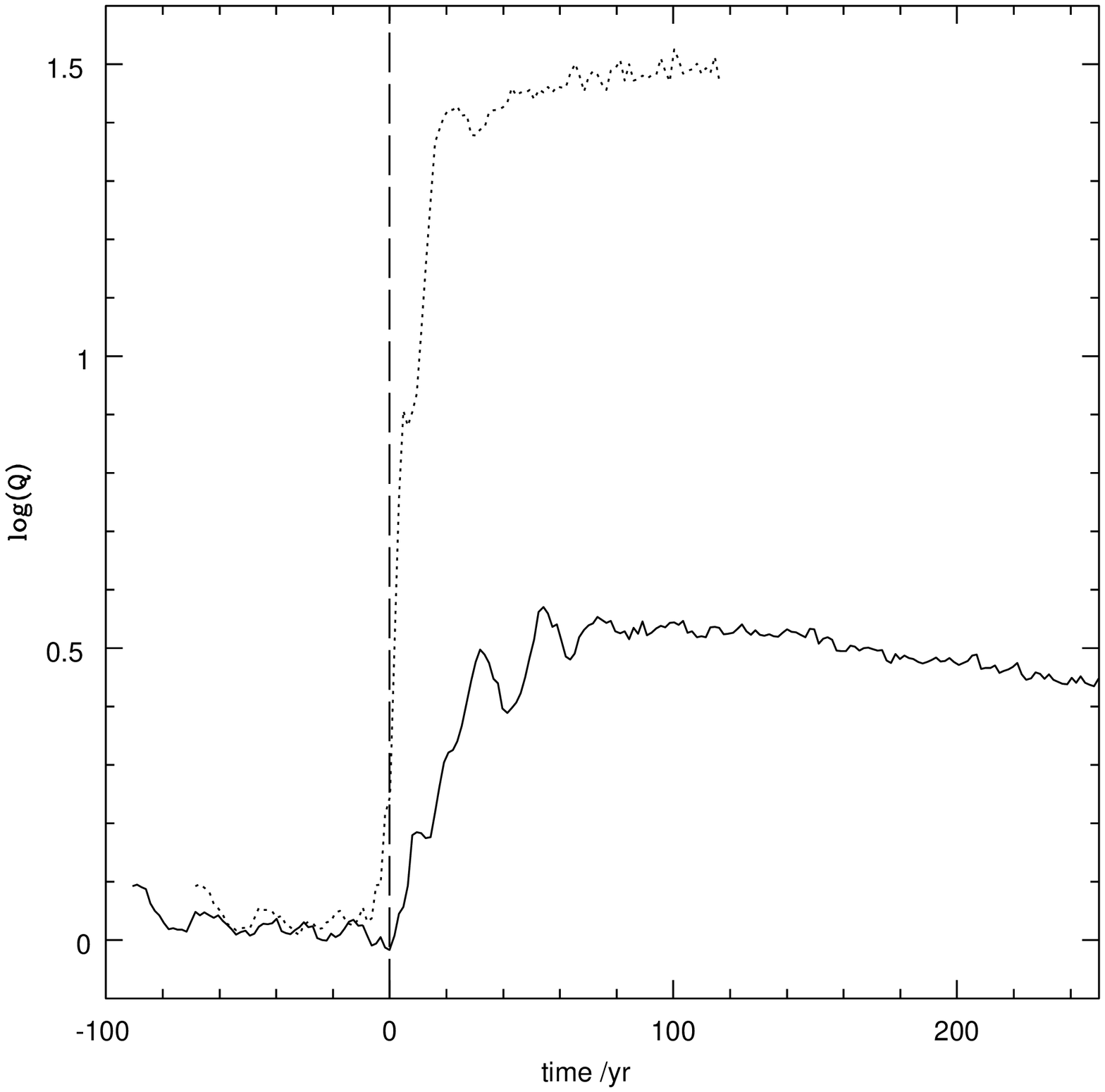,width=0.4\textwidth}}
\caption{\small  Azimuthal averages at
$R=8.5$AU of sound speed (left) and $\log(Q)$  (right) 
as a function of time
for simulations S1 (solid line; $M_{\rm pert}=0.1M_\odot$) and S2
(dotted line; $M_{\rm pert}=M_\odot$).}
\label{fig:massp}
\end{figure*}

\begin{figure}
\centerline{\epsfig{figure=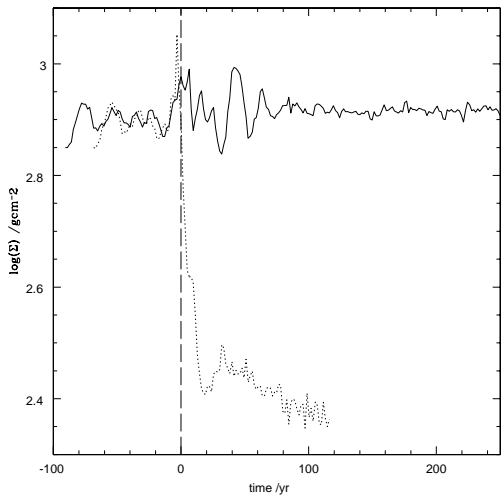,width=0.4\textwidth}}
\caption{\small  Azimuthal averages at
$R=8.5$AU of surface density $\Sigma$
as a function of time
for simulations S1 (solid line) and S2 (dotted
line; high perturber mass).}
\label{fig:sigmam}
\end{figure}

\subsection{Changing the perturber's mass}

Figure \ref{fig:massp} compares the Reference simulation with the
simulation with a perturbing star that is ten times heavier
(simulations S1 and S2). As expected, increasing the mass of the
perturber has a much larger effect on $Q$, and sound speed, consistent
with more vigorous shock heating. We note that $Q$ does not in this
case just track the evolution of the sound speed, implying that the
effect of the higher mass is to decrease the local column density by a
factor of a few. This is indeed confirmed in Fig. \ref{fig:sigmam},
that shows the time evolution of the disc surface density for the
reference case (solid line) and for simulation S2 (dotted line). While
in the reference case after the encounter the surface density is
roughly the same as the pre-encounter value, in the high perturber
mass case, it is decreased by roughly a factor 3.

\subsection{A hyperbolic encounter}

\begin{figure*}
\centerline{\epsfig{figure=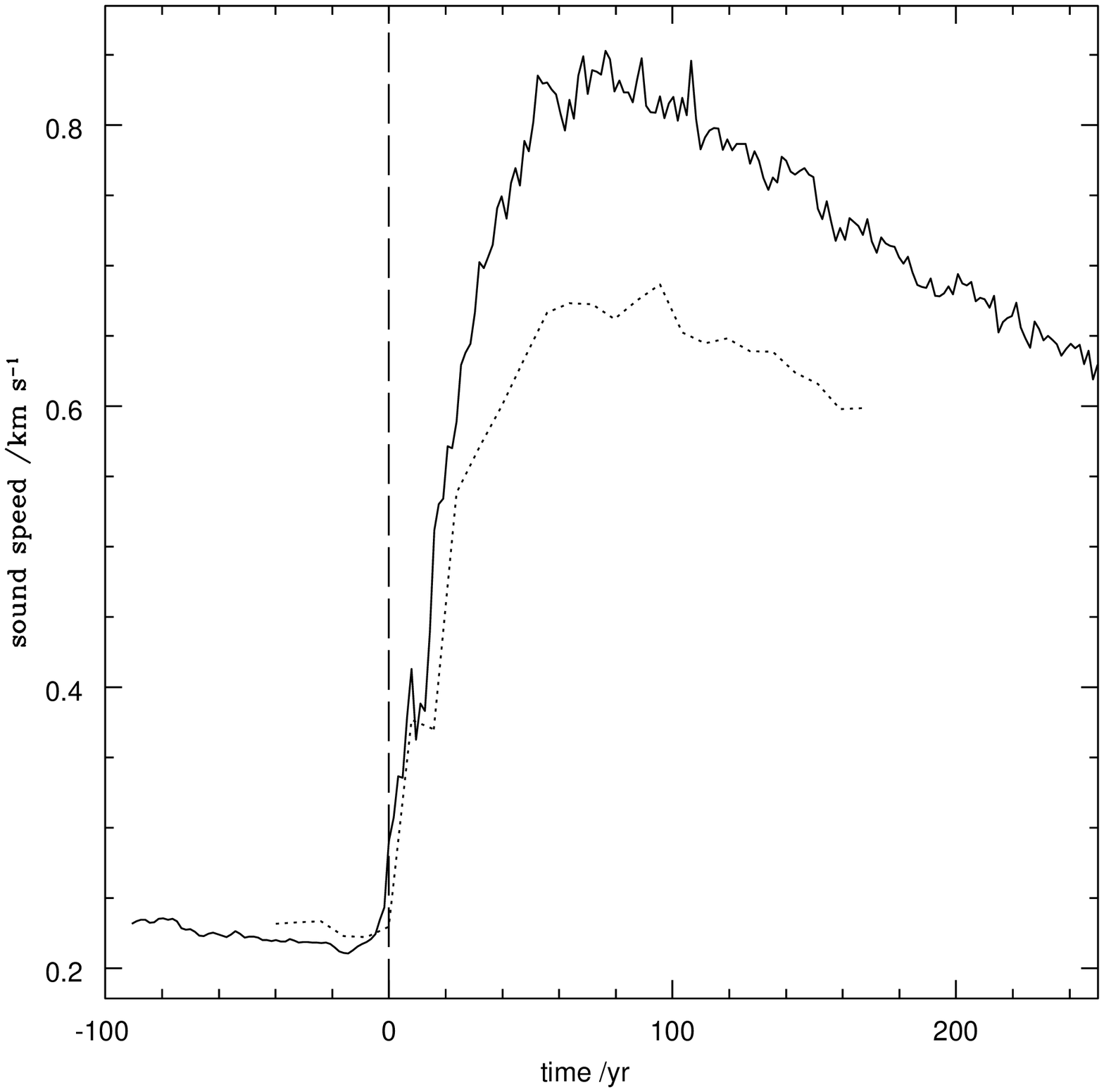,width=0.4\textwidth}
	    \epsfig{figure=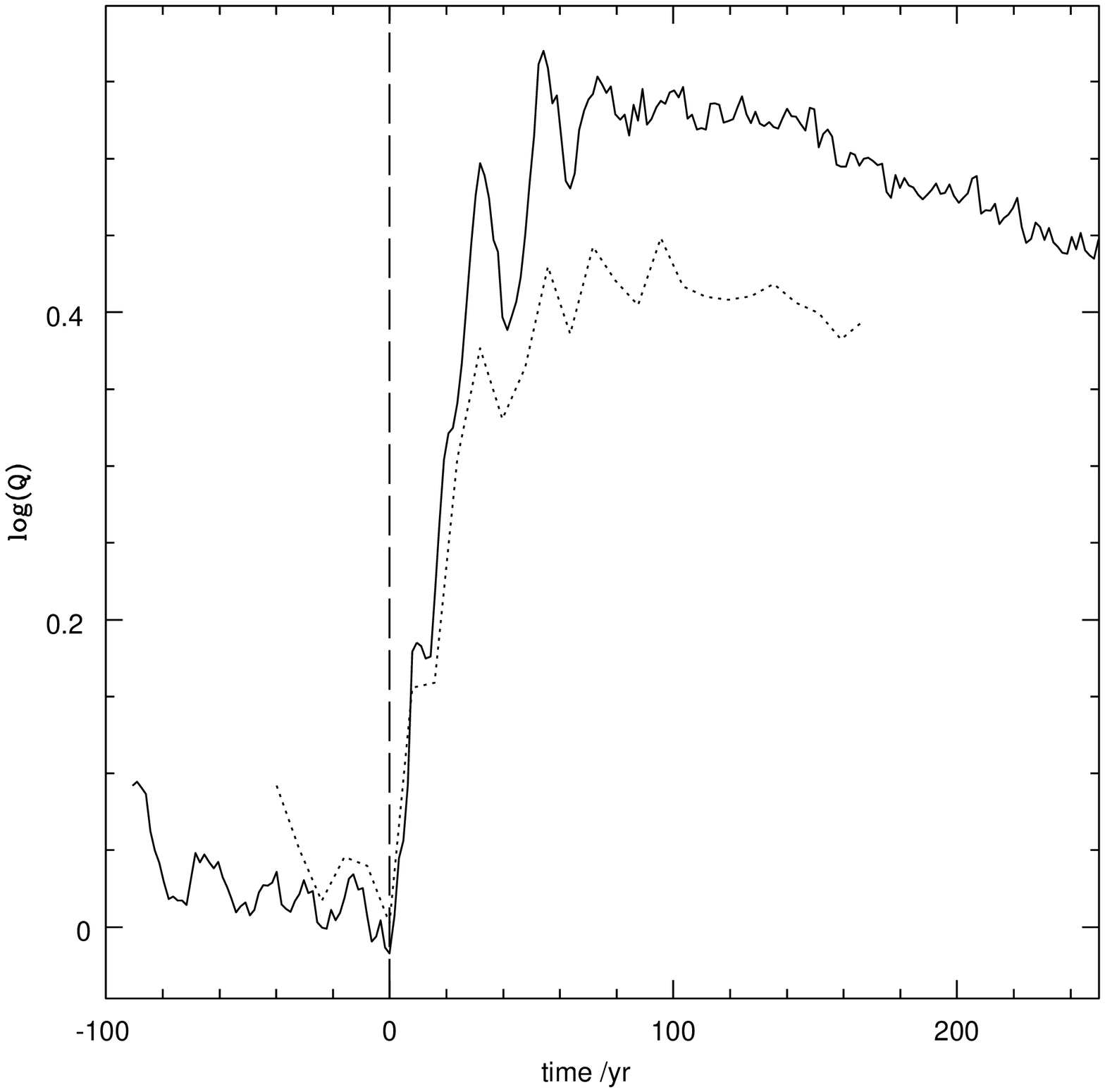,width=0.4\textwidth}}
\caption{\small  Azimuthal averages at
$R=8.5$AU of sound speed (left) and $\log(Q)$ (right) 
as a function of time
for simulations S1 (solid line; parabolic encounter) and S5 (dotted
line; hyperbolic encounter.}
\label{fig:ecc}
\end{figure*}

In Fig. \ref{fig:ecc} we show the results of a hyperbolic encounter
(simulation S5). In general, the effect is similar to the parabolic
one, but smaller in amplitude.  This is in general consistent with the
fact that the interaction occurs on a much faster timescale for a
hyperbolic encounter and is therefore less effective in perturbing the
disc structure.

\subsection{Changing the pericenter distance}

\begin{figure*}
\centerline{\epsfig{figure=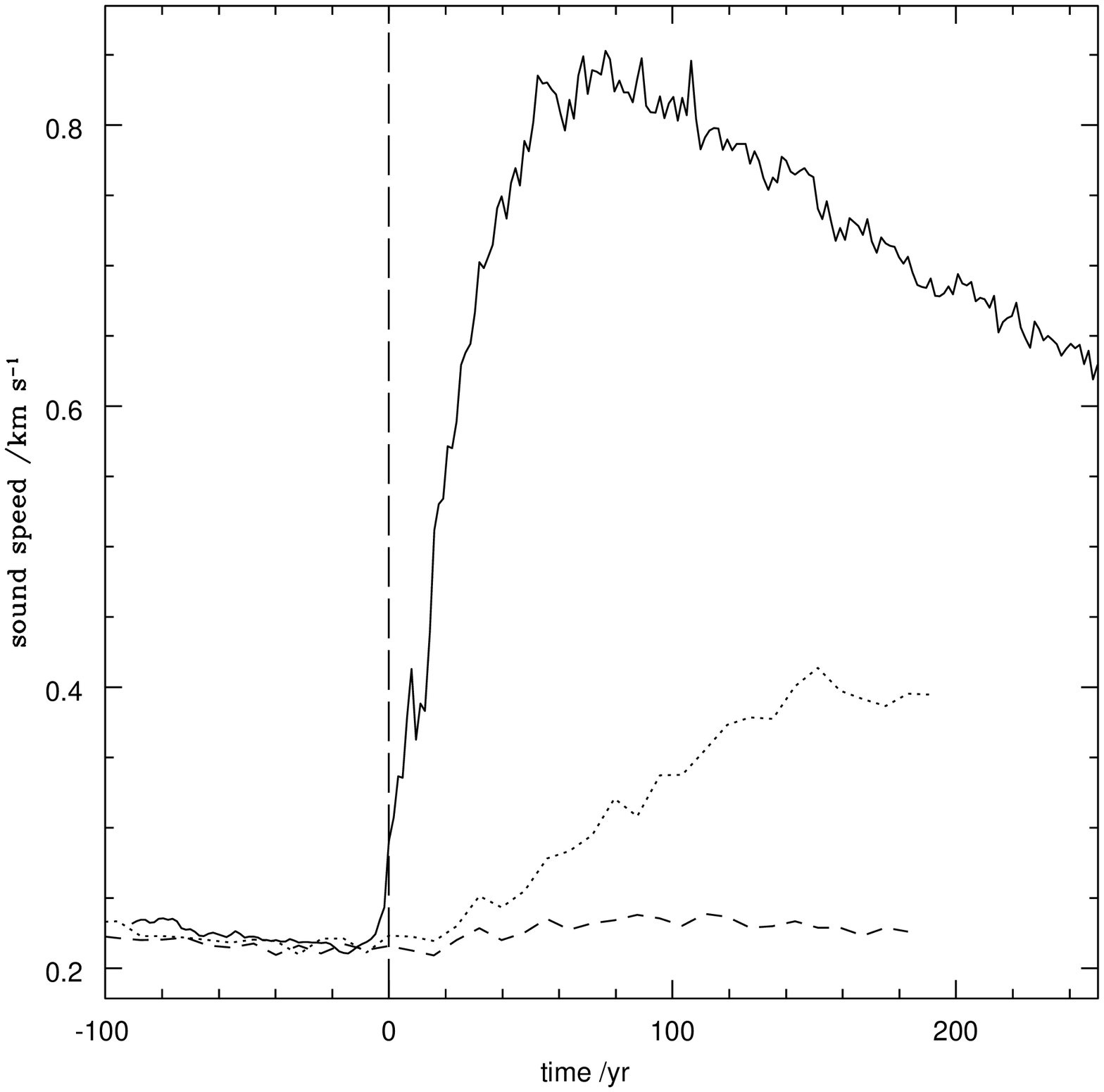,width=0.4\textwidth}
	    \epsfig{figure=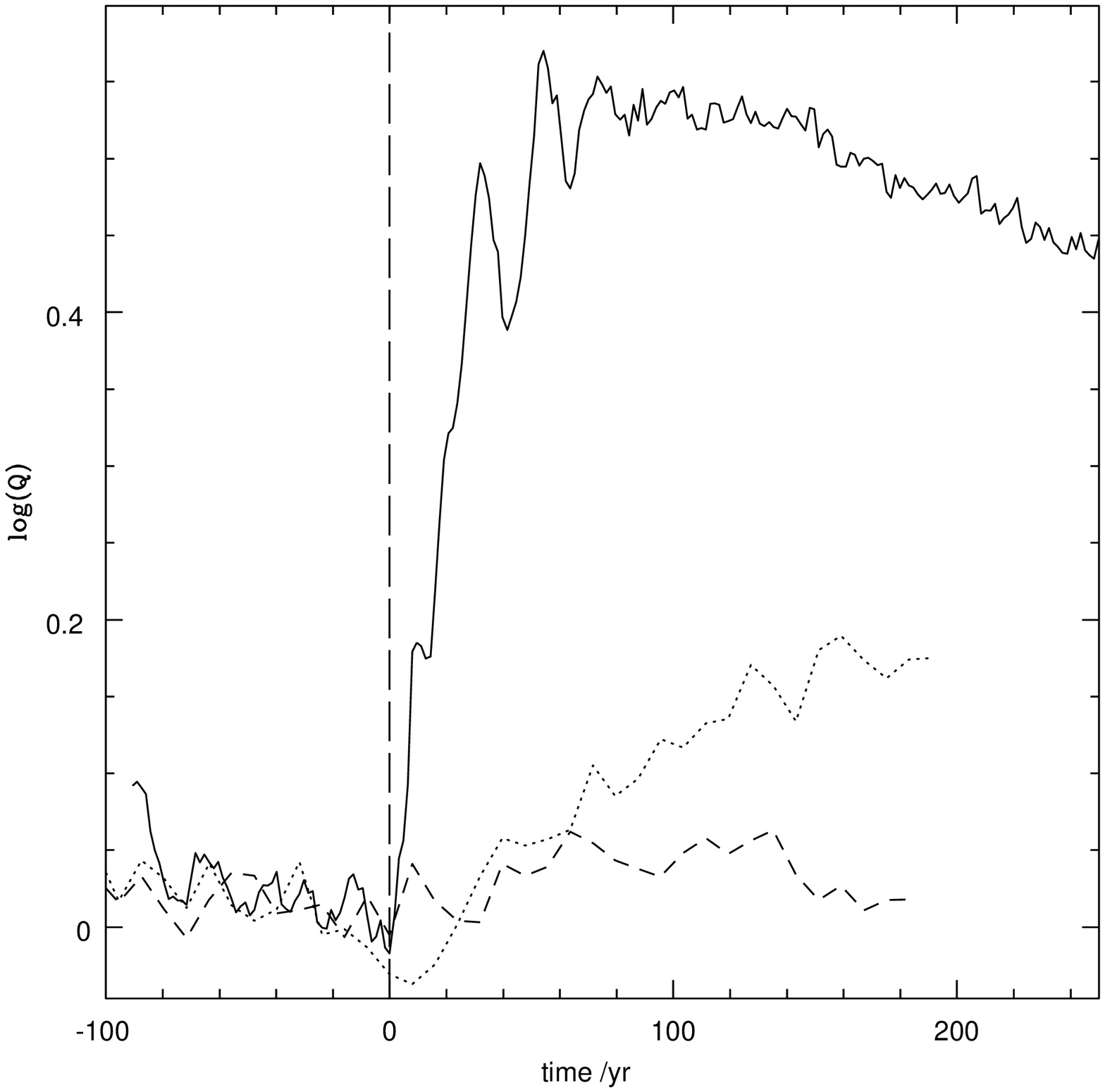,width=0.4\textwidth}}
          \caption{\small Azimuthal averages at $R=8.5$AU of sound
            speed (left) and $\log(Q)$ (right) as a function of time
            for simulations S1 (solid line; $R_{\rm peri} = 17$AU), S8
            (dotted line; $R_{\rm peri} = 30$AU) and S9 (dashed line;
            $R_{\rm peri} = 40$AU).}
\label{fig:peri}
\end{figure*}

Figure \ref{fig:peri} shows that overall, encounters with larger
pericenter distances have a much smaller effect on the temperature and
$Q$.  There is slight increase in temperature for the case where
$R_{\rm peri}=30$, whereas for the $R_{\rm peri}=40$ case also the
temperature is barely modified.

\subsection{A convergence test}

\begin{figure*}
\centerline{\epsfig{figure=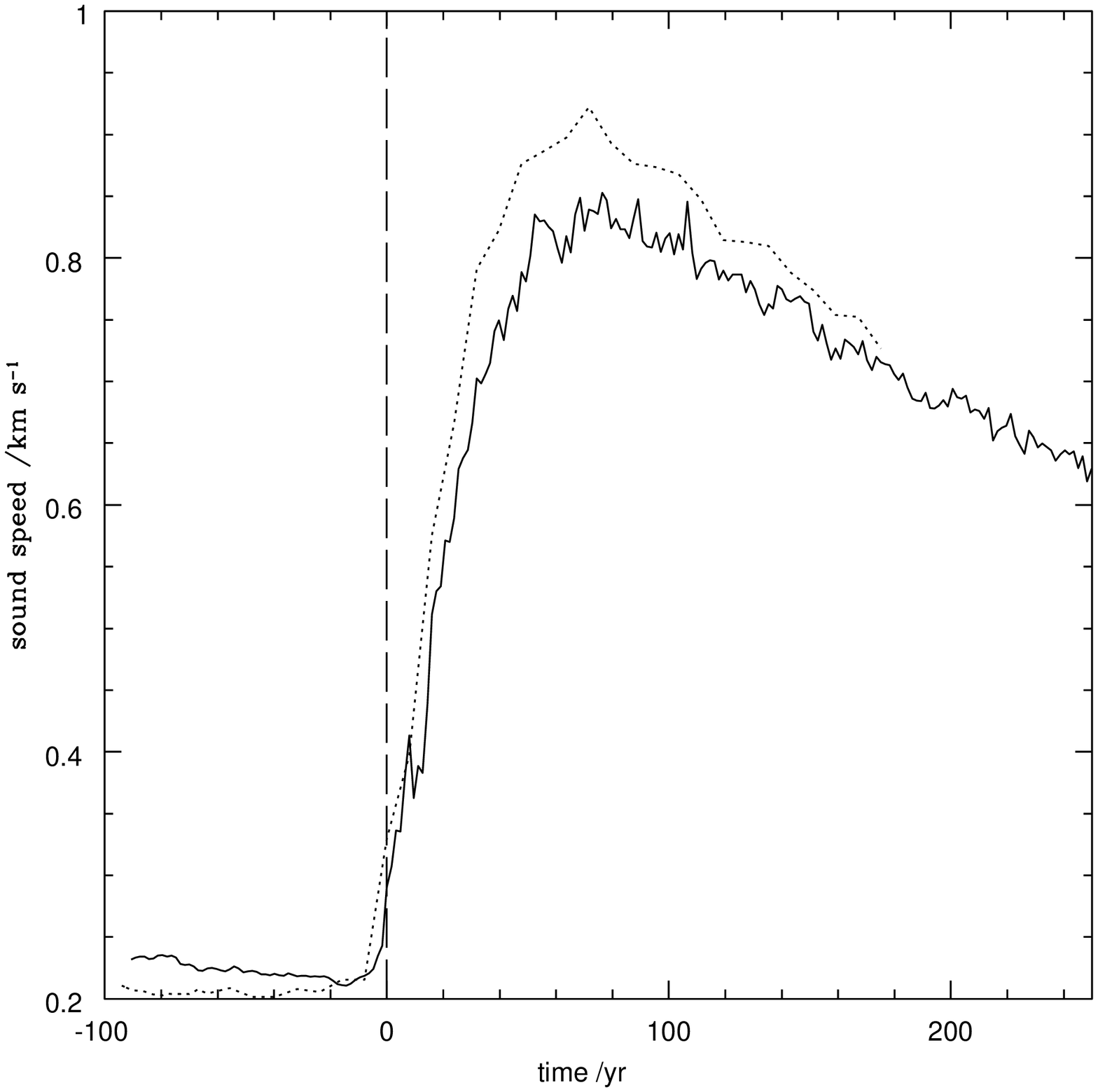,width=0.4\textwidth}
	    \epsfig{figure=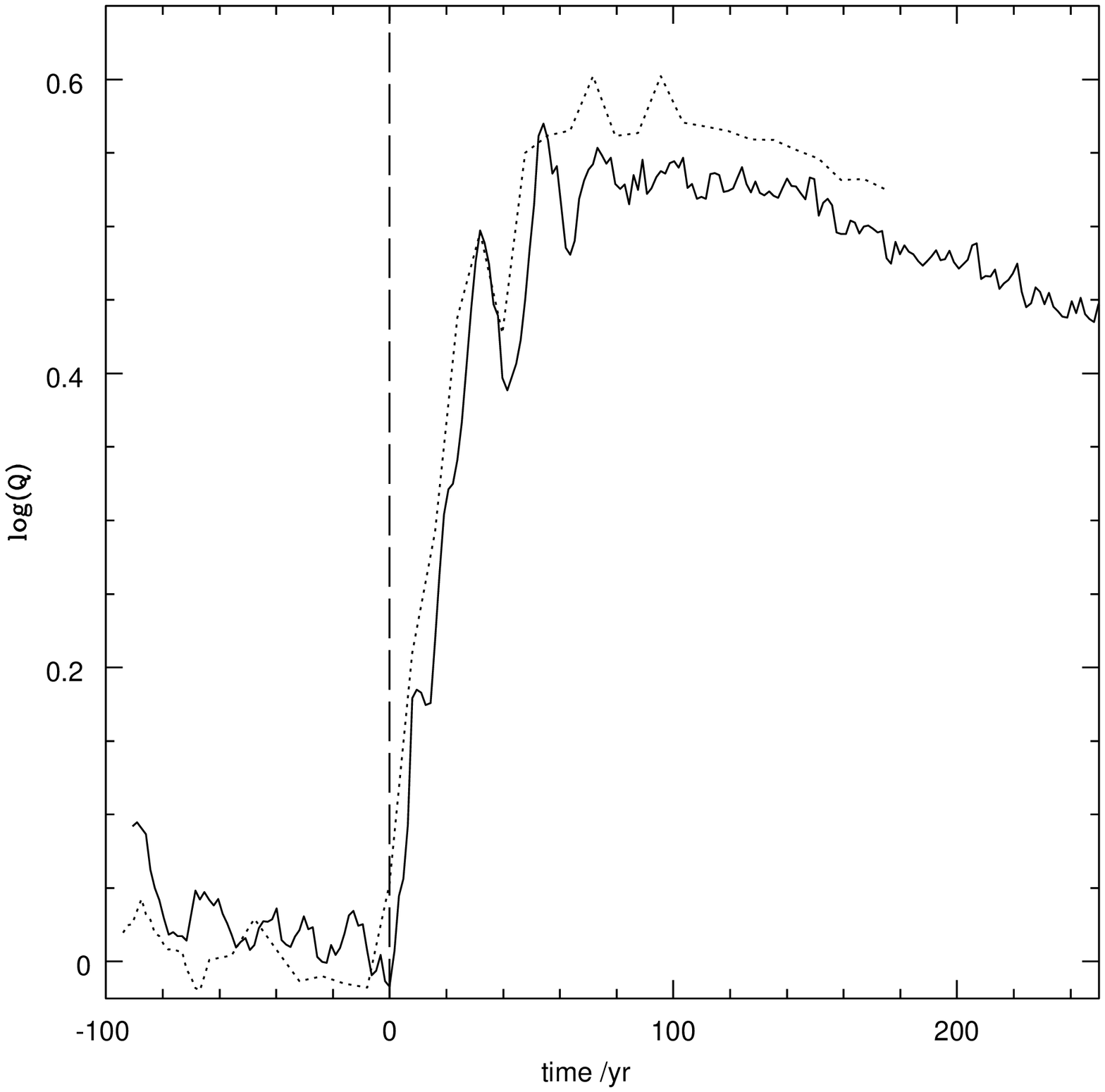,width=0.4\textwidth}}
          \caption{\small Azimuthal averages at $R=8.5$AU of sound
            spped (left) and $\log(Q)$ (right) as a function of time
            for simulations S1 (solid line; 250,000 particles) and H1
            (dotted line; 500,000 particles).}
\label{fig:res}
\end{figure*}

Finally, in Fig. \ref{fig:res} we show the results of our convergence
test. We have re-run simulation S1 at a higher resolution, by doubling
the number of SPH particles used. Figure ~\ref{fig:res} clearly shows
that there is little effect on the temperature and the Toomre
parameter when the number of particles is increased to 500,000.
Therefore, we can conclude that the resolution of the reference
simulation is adequate and does not affect the outcome of the results.

\section{Discussion and conclusions}
\label{sec:conclusion}

In this paper, we have analysed the effects of an encounter with a discless
companion on the structure and evolution of a protostellar disc. We have
improved on previous analyses by considering in more detail the energy balance
for the gas, thus going beyond the approximation of isothermal evolution used
in the past. This turns out to be very important, because the outcome of the
encounter is strongly dependent on the tidal heating induced in the disc.
Indeed, probably the main result of the present paper is the demonstration
that an encounter with a companion inhibits rather than promotes fragmentation
of a gravitationally unstable disc. This is in marked contrast with previous
results \citep{boffin98}, who instead have shown effective fragmentation
following an encounter. The main difference between our simulations and those
of \citet{boffin98} is that while \citet{boffin98} consider much larger discs,
for which the isothermal approximation is probably adequate, we instead follow
the heating and cooling processes in the disc and therefore allow the growth
of gravitational instabilities to feed back upon the thermodynamic state of
the disc. We include heating from $p\de V$ work and shocks and cool the disc
down with a cooling rate that is sufficiently small that the disc would not
fragment in isolation.

We thus conclude that the main parameter that determines the fragmentation of
discs is the cooling rate, regardless of the way in which the disc is driven
to instability (either through cooling, as in the simulations of
\citealt{gammie01,mayer02,rice03b}, or by a dynamical interaction, as
simulated here).  {\it A disc that does not fragment in isolation, cannot be
  driven to fragmentation through an interaction with a companion}. Indeed, if
anything, tidal heating could prevent fragmentation even in cases where the
disc would otherwise fragment, as shown by \citet{mayer04} (and previously,
but in two dimensions, by \citealt{nelson2000}), in the context of discs in a
binary system. \citet{boss06}, on the other hand, has shown that in binary
systems, disc fragmentation is indeed possible if the cooling timescale is
sufficiently small (estimated to be about $1-2$ orbital timescales by
\citealt{boss06}).  We note however, that the discs modelled by \citet{boss06}
were also shown to fragment even in the absence of binary companions. Here
again, it would therefore appear, that the cooling timescale is the main
determinant of whether discs fragment or not and that the role of companions
in either promoting or suppressing fragmentation is rather minor.

Our simulations adopt a simplified prescription for the cooling. In
reality, the cooling timescale is going to depend on disc properties,
such as density and opacity. In order to determine the stability of
discs against fragmentation it is therefore needed to carefully
evaluate the actual cooling rate. In the context considered here, it
is then interesting to note that \citet{whitworth06} have considered
this issue analytically and have shown that, with realistic opacity
laws, the effect of an encounter is to reduce the cooling rate,
further reinforcing the conclusions obtained in our work.

The simulations presented here involve an encounter between a
protostellar disc and a disc-less perturber. Our results are therefore
directly comparable to those of \citet{boffin98}. It is however also
interesting to consider the more complex behaviour and the much larger
parameter space involved in disc-disc interactions, as considered by
\citet{watkins98a,watkins98b}. We postpone this investigation to
future analyses.

\section*{Acknowledgements}

The simulations presented in this work have been performed at the UK
Astrophysical Fluid Facility (UKAFF).

\bibliographystyle{mn2e} 

\bibliography{lodato}

\begin{thebibliography}{}

\bibitem[\protect\citeauthoryear{{Bate}, {Bonnell} \& {Bromm}}{{Bate}
  et~al.}{2003}]{bate03b}
{Bate} M.~R.,  {Bonnell} I.~A.,    {Bromm} V.,  2003, MNRAS, 339, 577

\bibitem[\protect\citeauthoryear{{Bate} \& {Burkert}}{{Bate} \&
  {Burkert}}{1997}]{bate97}
{Bate} M.~R.,  {Burkert} A.,  1997, MNRAS, 288, 1060

\bibitem[\protect\citeauthoryear{Benz}{Benz}{1990}]{benz90}
Benz W.,  1990, in Buchler J.,  ed., The Numerical Modeling of Nonlinear
  Stellar Pulsations Kluwer, Dordrecht

\bibitem[\protect\citeauthoryear{{Boffin}, {Watkins}, {Bhattal}, {Francis} \&
  {Whitworth}}{{Boffin} et~al.}{1998}]{boffin98}
{Boffin} H.~M.~J.,  {Watkins} S.~J.,  {Bhattal} A.~S.,  {Francis} N.,
  {Whitworth} A.~P.,  1998, MNRAS, 300, 1189

\bibitem[\protect\citeauthoryear{Bonnell \& Bastien}{Bonnell \&
  Bastien}{1992}]{bonnell92}
Bonnell I.,  Bastien P.,  1992, ApJ, 401, 31

\bibitem[\protect\citeauthoryear{{Boss}}{{Boss}}{2006}]{boss06}
{Boss} A.~P.,  2006, ApJ, 641, 1148

\bibitem[\protect\citeauthoryear{{Clarke} \& {Pringle}}{{Clarke} \&
  {Pringle}}{1993}]{clarke93}
{Clarke} C.~J.,  {Pringle} J.~E.,  1993, MNRAS, 261, 190

\bibitem[\protect\citeauthoryear{Gammie}{Gammie}{2001}]{gammie01}
Gammie C.~F.,  2001, ApJ, 553, 174

\bibitem[\protect\citeauthoryear{{Goodwin}, {Kroupa}, {Goodman} \&
  {Burkert}}{{Goodwin} et~al.}{2006}]{goodwin06}
{Goodwin} S.~P.,  {Kroupa} P.,  {Goodman} A.,    {Burkert} A.,  2006, ArXiv
  Astrophysics e-prints

\bibitem[\protect\citeauthoryear{{Hall}, {Clarke} \& {Pringle}}{{Hall}
  et~al.}{1996}]{hall96}
{Hall} S.~M.,  {Clarke} C.~J.,    {Pringle} J.~E.,  1996, MNRAS, 278, 303

\bibitem[\protect\citeauthoryear{Lodato \& Rice}{Lodato \& Rice}{2004}]{LR04}
Lodato G.,  Rice W. K.~M.,  2004, MNRAS, 351, 630

\bibitem[\protect\citeauthoryear{Lodato \& Rice}{Lodato \& Rice}{2005}]{LR05}
Lodato G.,  Rice W. K.~M.,  2005, MNRAS, 358, 1489

\bibitem[\protect\citeauthoryear{Mayer et~al.,}{Mayer  et~al.}{2002}]{mayer02}
Mayer L.,  et~al., 2002, Science, 298, 1756

\bibitem[\protect\citeauthoryear{{Mayer}, {Quinn}, {Wadsley} \&
  {Stadel}}{{Mayer} et~al.}{2004}]{mayer04}
{Mayer} L.,  {Quinn} T.,  {Wadsley} J.,    {Stadel} J.,  2004, ApJ, 609, 1045

\bibitem[\protect\citeauthoryear{{Mayer}, {Wadsley}, {Quinn} \&
  {Stadel}}{{Mayer} et~al.}{2005}]{mayer05b}
{Mayer} L.,  {Wadsley} J.,  {Quinn} T.,    {Stadel} J.,  2005, MNRAS, 363, 641

\bibitem[\protect\citeauthoryear{Mejia, Durisen, Pickett \& Cai}{Mejia
  et~al.}{2005}]{mejia05}
Mejia A.~C.,  Durisen R.~H.,  Pickett M.~K.,    Cai K.,  2005, ApJ, 619, 1098

\bibitem[\protect\citeauthoryear{Monaghan}{Monaghan}{1992}]{monaghan92}
Monaghan J.~J.,  1992, ARA\&A, 30, 543

\bibitem[\protect\citeauthoryear{{Nelson}}{{Nelson}}{2000}]{nelson2000}
{Nelson} A.~F.,  2000, ApJ, 537, L65

\bibitem[\protect\citeauthoryear{Rice, Armitage, Bate, Bonnell, Jeffers \&
  Vine}{Rice et~al.}{2003}]{rice03b}
Rice W. K.~M.,  Armitage P.~J.,  Bate M.~R.,  Bonnell I.~A.,  Jeffers S.~V.,
  Vine S.~G.,  2003, MNRAS, 346, L36

\bibitem[\protect\citeauthoryear{{Rice}, {Lodato} \& {Armitage}}{{Rice}
  et~al.}{2005}]{RLA05}
{Rice} W.~K.~M.,  {Lodato} G.,    {Armitage} P.~J.,  2005, MNRAS, 364, L56

\bibitem[\protect\citeauthoryear{{Scally} \& {Clarke}}{{Scally} \&
  {Clarke}}{2001}]{SC01}
{Scally} A.,  {Clarke} C.,  2001, MNRAS, 325, 449

\bibitem[\protect\citeauthoryear{{Watkins}, {Bhattal}, {Boffin}, {Francis} \&
  {Whitworth}}{{Watkins} et~al.}{1998a}]{watkins98a}
{Watkins} S.~J.,  {Bhattal} A.~S.,  {Boffin} H.~M.~J.,  {Francis} N.,
  {Whitworth} A.~P.,  1998a, MNRAS, 300, 1205

\bibitem[\protect\citeauthoryear{{Watkins}, {Bhattal}, {Boffin}, {Francis} \&
  {Whitworth}}{{Watkins} et~al.}{1998b}]{watkins98b}
{Watkins} S.~J.,  {Bhattal} A.~S.,  {Boffin} H.~M.~J.,  {Francis} N.,
  {Whitworth} A.~P.,  1998b, MNRAS, 300, 1214

\bibitem[\protect\citeauthoryear{{Whitworth}, {Bate}, {Nordlund}, {Reipurth} \&
  {Zinnecker}}{{Whitworth} et~al.}{2006}]{whitworth06}
{Whitworth} A.,  {Bate} M.~R.,  {Nordlund} A.,  {Reipurth} B.,    {Zinnecker}
  H.,  2006, ArXiv Astrophysics e-prints

\end{thebibliography}

\end{document}